\documentclass[pra, aps, twocolumn, groupedaddress, showpacs, superscriptaddress]{revtex4-2}
\usepackage{amsmath,amssymb,amsthm}
\usepackage{algorithm}
\usepackage{hyperref}
\usepackage{titlesec}
\usepackage{subfigure}
\usepackage{tikz}
\usetikzlibrary{quantikz}
\usepackage{xcolor}
\usepackage{color}
\usepackage{hyperref}
\usepackage{orcidlink}
\usepackage{graphicx}
\usepackage{listings}
\usepackage{xcolor}
\usepackage{xspace}
\usepackage{algorithm}
\usepackage{algorithmicx}
\usepackage{algpseudocode}
\usepackage{float}

\providecommand{\openone}{\leavevmode\hbox{\small1\kern-4.3pt\normalsize1}}

\theoremstyle{plain}

\theoremstyle{definition}

\begin{document}

\title{Optimizing Multi-Hop Quantum Communication using Bidirectional Quantum Teleportation Protocol}

\author{N. Ikken \orcidlink{0009-0004-4513-5977}}\affiliation{LPHE-Modeling and Simulation, Faculty of Sciences, Mohammed V University in Rabat, Morocco.}
\author{P. Kumar \orcidlink{0009-0003-7422-2087}}\affiliation{Department of Computer Science and Information Engineering, National Taiwan University of Science and Technology, Taipei, Taiwan.}
\author{A. Slaoui \orcidlink{0000-0002-5284-3240}}\affiliation{LPHE-Modeling and Simulation, Faculty of Sciences, Mohammed V University in Rabat, Morocco.}\affiliation{Centre of Physics and Mathematics, CPM, Faculty of Sciences, Mohammed V University in Rabat, Rabat, Morocco.}\affiliation{EIAS Data Science Lab, College of Computer and Information Sciences, and Center of Excellence in Quantum and Intelligent Computing, Prince Sultan University, Riyadh 11586, Saudi Arabia.}
\author{B. Kar \orcidlink{0000-0003-0534-6652}}\affiliation{Department of Computer Science and Information Engineering, National Taiwan University of Science and Technology, Taipei, Taiwan.}
\author{R. Ahl Laamara \orcidlink{0000-0002-8254-9085}}\affiliation{LPHE-Modeling and Simulation, Faculty of Sciences, Mohammed V University in Rabat, Morocco.}\affiliation{Centre of Physics and Mathematics, CPM, Faculty of Sciences, Mohammed V University in Rabat, Rabat, Morocco.}
\author{M. Almousa}\affiliation{Department of Information Technology, College of Computer and Information Sciences, Princess Nourah bint Abdulrahman University, Riyadh 11671, Saudi Arabia.}
\author{A. A Abd El-Latif \orcidlink{0000-0002-5068-2033}}\affiliation{EIAS Data Science Lab, College of Computer and Information Sciences, and Center of Excellence in Quantum and Intelligent Computing, Prince Sultan University, Riyadh 11586, Saudi Arabia.}\affiliation{Department of Mathematics and Computer Science, Faculty of Science, Menoufia University, Shebin El-Koom 32511, Egypt.}

\begin{abstract}
In this paper, we introduce a new method for Bidirectional Quantum Teleportation called Bidirectional Quantum Teleportation using the Modified Dijkstra Algorithm and Quantum Walk (BQT-MDQW). This method uses different types of entangled states, such as the GHZ-Bell state, W-Bell state, and Cluster-Bell state, to improve quantum communication in multi-hop quantum wireless networks. We focus on the W-Bell state and compare the quantum Dijkstra algorithm with the classical Dijkstra method to see which one works better. We apply both versions to quantum and classical simulators, measuring their performance through fidelity, memory utilization, and throughput calculations. Our results show that the shortest path problem may be solved with significantly reduced computer complexity using the quantum Dijkstra algorithm based on quantum walks. The introduction of a quantum walk, which permits dynamic transitions between quantum channels and the effective exploration of quantum network states, is an important part of the protocol. Using the capacity of the quantum walk to adjust to changing quantum states, we also introduce a method for successfully identifying unitary matrices under varying quantum channels. The bidirectional teleportation structure of the protocol is designed to solve the multi-hop teleportation problem in quantum wireless networks. In addition, we present quantum Dijkstra's algorithm, which uses quantum gates to significantly decrease computational complexity and solve the networking problem by building on the quantum walk framework. This method shows how quantum computing may be used to solve arbitrary optimization issues such as the shortest path problem. Finally, we present a novel multi-hop quantum teleportation system encompassing both unidirectional and bidirectional communication, as introduced in the quantum Dijkstra algorithm system. This system significantly enhances communication in quantum networks by enabling efficient and reliable information transfer across multiple nodes. Moreover, we demonstrate how these quantum teleportation protocols seamlessly integrate with entangled states, including W-Bell states, GHZ-Bell states, and Cluster-Bell states, further improving network performance and resilience. Through extensive simulations and testbed experiments, we analyze the behavior of these states within quantum networks, highlighting their impact on optimization, routing efficiency, and overall network robustness.
\par
\vspace{0.25cm}
\end{abstract}

\maketitle
\section{Introduction}
Using shared entanglement and classical communication, quantum teleportation allows Alice to send an unknown qubit state to Bob, who is located far away \cite{1993}. Bennett et al. published the first quantum teleportation protocol in 1993, which was later shown to be experimentally successful \cite{1993,1997}. The use of quantum teleportation has since been expanded to multi-qubit and higher-dimensional systems \cite{2008,24}. Instead of a one-way transfer, researchers have created a bidirectional teleportation mechanism that allows Alice and Bob to exchange qubits simultaneously \cite{8,9}. The shortest path problem has been extensively studied in computer science and is a basic topic in graph theory. To solve this problem, many classical algorithms have been developed, the most well-known and often used of which is Dijkstra's algorithm. Edsger W. Dijkstra first presented the algorithm in 1959 \cite{dijk}. It begins at a specified source node and proceeds carefully through the network. The node with the least known distance from the source among those that have not been decided upon is chosen at each iteration, after which it updates the distances of its closest neighbors. Dijkstra's algorithm can effectively find the shortest pathways from the source to every other node in the graph, thanks to this iterative process of expanding the frontier and updating probable distances. Unidirectional quantum teleportation (UQT) allows the transfer of an unknown quantum state from Alice to Bob in a single direction, employing entanglement and classical communication \cite{Zurek}. This process enables the perfect reconstruction of a quantum state without physically moving the qubit. Since the measurement outcome is random, Alice sends two classical bits to Bob to indicate which Bell states she observed. Bob then applies a specific unitary transformation (Pauli operations) \cite{Chen} based on Alice’s message to recover the exact quantum state. This method also consumes entanglement and requires classical communication. Unidirectional teleportation is crucial for quantum networks, distributed computing, and cryptographic protocols \cite{Elliott}, ensuring secure and efficient quantum state transfer.

In Bidirectional Quantum Teleportation (BQT), entanglement is essential. Still, there is an alternate method in which initial state-to-state entanglement is created during the BQT teleportation process rather than being necessary \cite{10, BQT}. A method called a quantum walk, which is the quantum walk of a classical random walk \cite{11}, can accomplish this, especially when a specially created quantum walk is used \cite{12}. Graph theory, quantum information processing, and quantum simulation all use quantum walks, which are essential to quantum computing. Their computational approach has shown great promise in recent years \cite{13,14}. Several approaches to qubit teleportation make use of the ideas of quantum walks \cite{15,16}.

The problem of finding the shortest route between two nodes in a network is a fundamental challenge in computer science, with applications that span communication networks \cite{Yurke}, transportation, robotics, and more. Classical algorithms, such as the Dijkstra algorithm, have been extensively used to solve this problem efficiently. However, as networks grow in size and complexity, classical approaches face limitations in computational efficiency, particularly when handling large-scale, dynamic, or uncertain environments.

Recent developments in quantum information science have further refined the theoretical models and practical methods that enable quantum communication and teleportation, building on previous discussions \cite{maj}. Quantum channels are now much more reliable because of the introduction of quantum error correction techniques and enhanced entanglement distribution strategies, opening the way for real-world quantum network applications. Examining hybrid quantum systems, which combine discrete and continuous variables, has also created new opportunities for improved security in quantum communication and scalable quantum computing \cite{min}. As environmental interactions continue to be an important barrier to preserving quantum coherence, new developments in metrological applications and noise-resistant teleportation are going to influence the direction of quantum technology \cite{bet}. All of these advancements support the continuous endeavor to connect basic quantum theories with their application in complicated quantum systems \cite{een}.

Determining the Multi-criteria Decision Making (MCDM) shortest path is one of the basic optimization issues that quantum computing has shown promise to solve in recent years\cite{17}. In some situations, quantum algorithms may perform better than their conventional equivalents employing concepts such as quantum superposition, entanglement, and interference\cite{18}. The main obstacle in creating practical quantum shortest-path algorithms is the limitations of quantum hardware, such as the limited quantity of qubits that may be used and high error rates. These limitations impact quantum algorithms' reliability in everyday applications and make it difficult to scale them to larger and more complicated issue situations. In addition, there is still work to be done in the field to create quantum algorithms that can handle dynamic graph structures and take into account a variety of actual limitations.
Bidirectional Quantum Teleportation using the Modified Dijkstra Algorithm and Quantum Walk (BQT-MDQW), we present an optimal method for Bidirectional Quantum Teleportation in this paper. Using quantum walk dynamics to analyze several paths at once and a modified Dijkstra algorithm to identify the fastest and most reliable path between nodes, this technique improves the effectiveness of quantum information transfer. By combining these methods, we hope to reduce resource overhead and maximize fidelity in complex quantum networks, increasing the possibility of bidirectional Quantum Teleportation for the next quantum communication systems.

In conclusion, this work highlights the revolutionary possibilities of quantum computing in solving combinatorial optimization issues and provides a comprehensive evaluation of modern quantum shortest-path methods. In this work:

\begin{itemize}
    \item {We examine how quantum in communication, the proposed algorithm BQT-MDQW can be used to solve the shortest-path problem in UQT and BQT.}
    \item {We begin by exploring the fundamental concepts of quantum computing and the entangled channel we have chosen to utilize, highlighting its potential in solving combinatorial optimization problems.}
    \item {We conduct both simulation-based and circuit design-based experiments on a quantum computer, analyzing various entangled states. Our study explores the behavior of quantum states in different scenarios, providing insight into their dynamics within the framework of quantum communication.}
\end{itemize}

\section{Multihop Bidirectional Quantum Teleportation}  
\begin{figure}[h]
    \centering
    \includegraphics[width=0.5\textwidth]{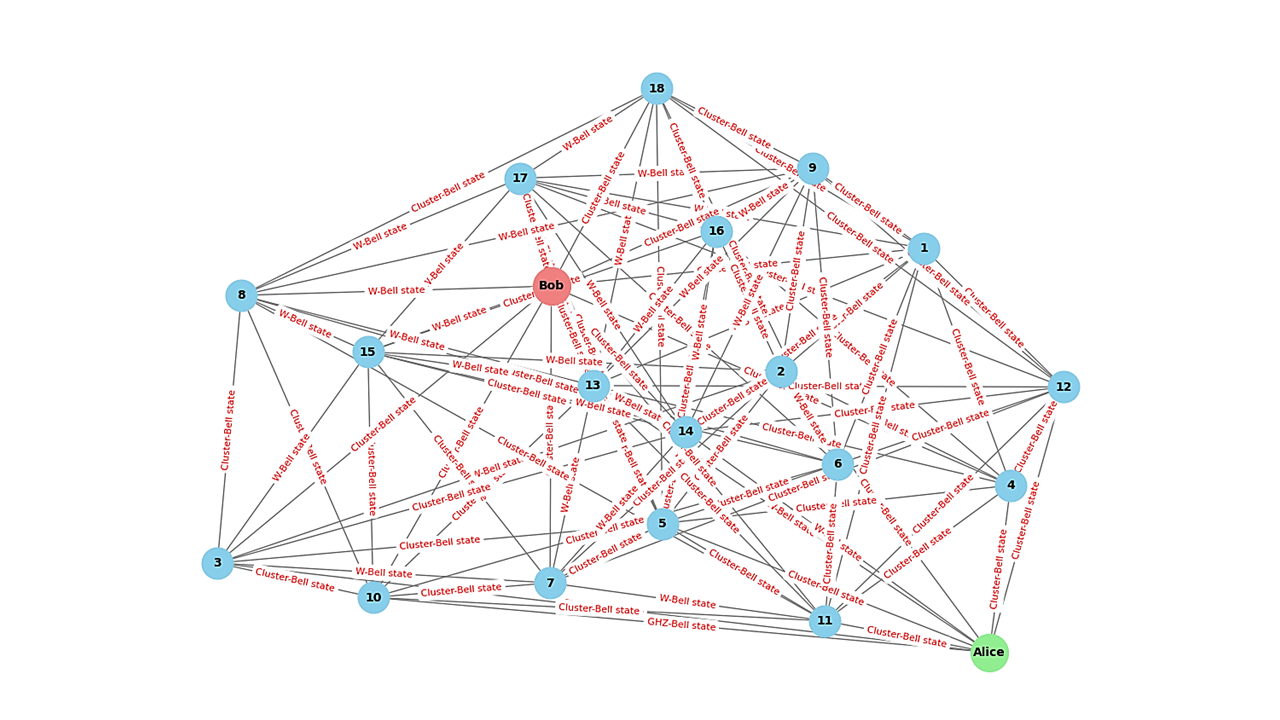}
    \caption{Schematic representation of multi-node quantum entanglement teleportation network}
    \label{fig1}
\end{figure}

Ref.\cite{Winterbone2015} describes the procedure as equivalent to standard quantum teleportation in the case of one-hop quantum communication. However, in order to relay quantum information over longer distances, the process becomes more complicated for multihop quantum communication with 2. This requires numerous intermediate nodes. In multihop quantum teleportation, as shown in Fig. \ref{fig1}, the sender and receiver are coupled by two different entangled resources each time: a 4-qubit cluster-Bell state and a GHZ-Bell state, as in the structures presented in Refs.\cite{2013,2014}. 

Alice has an unknown state:
\begin{align}
    |\phi\rangle_A = \alpha|0\rangle_A + \beta|1\rangle_A,
\end{align}

Where the coefficients $\alpha$ and $\beta$ are determined by the parametrization:
\begin{align}
    \alpha = \cos\left(\frac{\theta_A}{4}\right), \quad \beta = \sin\left(\frac{\theta_A}{4}\right),
\end{align}

with $\theta_A$ encoding the phase-space structure of her state in the Bloch sphere representation. Similarly, Bob's quantum state is expressed as
\begin{align}
    |\phi\rangle_B = \gamma|0\rangle_B + \eta|1\rangle_B,
\end{align}
where the amplitudes $\gamma$ and $\eta$ are given by
\begin{align}
    \gamma = \cos\left(\frac{\theta_B}{4}\right), \quad \eta = \sin\left(\frac{\theta_B}{4}\right),
\end{align}
with $\theta_B$ characterizing the orientation of Bob's qubit state within his Bloch sphere.\par

These parameterized versions illustrate the Bloch sphere's fundamental role in characterizing single-qubit quantum states. The geometric dependence shows the interaction between the probability amplitudes in the computational basis $\{|0\rangle, |1\rangle\}$ and their geometric representation, and it shows the continuous symmetry of the qubit states under rotations. 
\subsection{Bidirectional Quantum Teleportation using a Werner-Bell State}\label{12}
In \cite{2006}, Agrawal et al. showed that there is a special class of W states, which is denoted as
\begin{multline}
    |W_n\rangle = \frac{1}{\sqrt{2+2n}} \left( |100\rangle + \sqrt{n} e^{i\beta}+ \sqrt{n+1}e^{i\eta}|001\rangle \right)
\end{multline}
where $n$ is a real number, $\beta$ and $\eta$ are phases. Note that when $n = 1$, $\beta = \eta = 0$, $|W_n\rangle$ is reduced to $|W_n\rangle = \frac{1}{2} \left( |100\rangle + |010\rangle + \sqrt{2}|001\rangle  \right)_{123}$. To begin, we assume that Alice and Bob share an entangled composite state, referred to as the $W$-$Bell$ state, which is given by
\begin{multline}
    |\omega\rangle_{12345} = \frac{1}{2} \left( |100\rangle + |010\rangle + \sqrt{2}|001\rangle  \right)_{123} \otimes \frac{1}{\sqrt{2}} \left( |00\rangle + |11\rangle \right)_{45}\\ 
    = \frac{1}{2\sqrt{2}} ( |10000\rangle + |01000\rangle +  \sqrt{2}|00100\rangle \\+ |10011\rangle + |01011\rangle + \sqrt{2}|00111\rangle )_{12345}
\end{multline}

In this state, Bob holds particles 2, 3, and 4, while Alice holds particles 1 and 5. This bipartite distribution of subsystems serves as the basis for the quantum communication of the two parties and reflects their spatial separation. The initial state of the entire system, including Alice's and Bob's local qubits, can then be expressed as
\begin{align}
    |\Omega\rangle_{12345AB} = |\omega\rangle_{12345} \otimes |\phi\rangle_A \otimes |\phi\rangle_B,
\end{align}
where $|\phi\rangle_A$ and $|\phi\rangle_B$ represent the unknown quantum states of Alice's and Bob's qubits, respectively. As shown in the figure, this total system state forms the starting point for analyzing entanglement-based quantum communication protocols.
\begin{widetext}

\begin{figure}[h]
    \centering
    \includegraphics[width=1\textwidth]{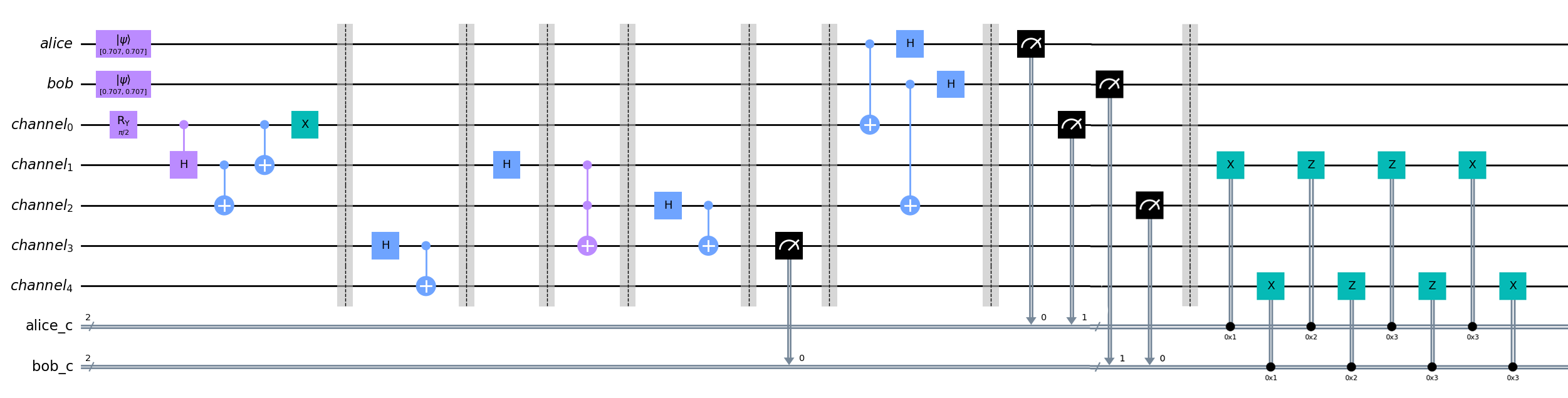}
    \caption{Quantum circuit for bidirectional quantum teleportation using a W-Bell state as the shared entangled resource.}
    \label{fig2}
\end{figure}
\end{widetext}

The circuit shown in Fig.\ref{fig2} for bidirectional quantum teleportation is designed to use entanglement as a resource distributed across multiple nodes. Entanglement is a resource that is dispersed over several nodes in the circuit for bidirectional quantum teleportation. A Hadamard gate and Controlled-NOT (CNOT) gates are used to construct the W state, $\frac{1}{\sqrt{2}}(|000\rangle + |111\rangle)$, which ensures maximal entanglement for the teleportation protocol. Additional qubits are created in a Bell state, $\frac{1}{\sqrt{2}}(|00\rangle + |11\rangle)$, and entangled with the $W$ state using additional CNOT operations to try to allow multiparty entanglement. Alice and Bob then distribute these entangled qubits across the network. Both parties collapse the entangled state and project the remaining qubits into a correlated state by performing local Bell-State Measurements (BSM) on their respective particles and unknown quantum states, $|\phi\rangle_A$ and $|\phi\rangle_B$. Since the measurement results are sent across classical communication channels, the original quantum states on the corresponding target qubits can be recovered by applying unitary corrections using Pauli operators (gates X and Z). The successful transfer of quantum states across numerous nodes is ensured by applying this procedure continuously at each network hop.

To perform the calculations step by step as described, we proceed as follows:

\subsection{Calculation Steps for Bidirectional Quantum Teleportation}

\emph{\textbf{Step 1: Applying the Hadamard Gate to qubit 2.}}
Bob applies a Hadamard gate ($H$) to qubit 2 of the $W$ state. The state evolves as follows:
\begin{multline}
    |\Omega\rangle_{12345} = \\ \frac{1}{\sqrt{2}} \big(|100\rangle + |010\rangle + |001\rangle\big)_{123} \otimes \frac{1}{\sqrt{2}} \big(|00\rangle + |11\rangle\big)_{45} \\
    \xrightarrow{H_2} \frac{1}{2\sqrt{2}} \big[|10\rangle_{13}(|0\rangle + |1\rangle)_2 + |00\rangle_{13}(|0\rangle - |1\rangle)_2 \\ + 
    \sqrt{2}|01\rangle_{13}(|0\rangle + |1\rangle)_2\big] \otimes \big(|00\rangle + |11\rangle\big)_{45}.
\end{multline}

\emph{\textbf{Step 2: Applying the Controlled-Controlled-NOT (CCNOT) Gate.}} 
Bob performs a CCNOT gate on qubit 2, with qubit 2 as a control qubit, another control qubit in state $|1\rangle$ and a target qubit in state $|0\rangle$. If qubit 2 is in the state $|0\rangle$, the target qubit remains unchanged. If qubit 2 is in the state $|1\rangle$, the target qubit flips. Assuming the outcome of the target qubit measurement is $|0\rangle$, qubit 2 remains in the state $|0\rangle$. The state becomes:
\begin{multline}
    |\omega'\rangle_{12345} \xrightarrow{CCNOT} \frac{1}{4\sqrt{2}} \big[|10000\rangle + |11000\rangle + |00000\rangle \\ -
    |01000\rangle + \sqrt{2}|00100\rangle + \sqrt{2}|01110\rangle+|10011\rangle \\ + |11011\rangle + |00011\rangle - |01011\rangle + \sqrt{2}|00111\rangle + \sqrt{2}|01101\rangle\big]_{12345}
\end{multline}

\emph{\textbf{Step 3: Applying the Hadamard and CNOT gate.}}
Bob applies a Hadamard to qubit 3 and CNOT gate to qubits 3 and 4, with qubit 3 as the control qubit and particle 4 as the target qubit. The state evolves as follows:
\begin{multline}
    |\omega"\rangle_{12345} \xrightarrow{H_{3}} \\ \frac{1}{8} \big[|1000\rangle_{1245}(|0\rangle + |1\rangle)_{3} + |1100\rangle_{1245}(|0\rangle + |1\rangle)_{3} \\ +|0000\rangle_{1245}(|0\rangle - |1\rangle)_{3} - |0100\rangle_{1245}(|0\rangle - |1\rangle)_{3} \\ +  \sqrt{2}|0000\rangle_{1245}(|0\rangle - |1\rangle)_{3} + \sqrt{2}|0110\rangle_{1245}(|0\rangle - |1\rangle)_{3} \\ +  |1011\rangle_{1245}(|0\rangle + |1\rangle)_{3} + |1111\rangle_{1245}(|0\rangle + |1\rangle)_{3} \\ +  |0011\rangle_{1245}(|0\rangle + |1\rangle)_{3} - |0111\rangle_{1245}(|0\rangle + |1\rangle)_{3} \\ +  \sqrt{2}|0011\rangle_{1245}(|0\rangle - |1\rangle)_{3} +  \sqrt{2}|0101\rangle_{1245}(|0\rangle - |1\rangle)_{3}\big].
\end{multline}

Now we apply the CNOT gate:
\begin{multline}
    |\omega'''\rangle_{12345} \xrightarrow{CNOT_{3,4}} \frac{1}{8\sqrt{2}} \big[|10000\rangle_{12345} + |10110\rangle_{12345} \\ + |11000\rangle_{12345}+ |11110\rangle_{12345}+|00000\rangle_{12345}+ |00110\rangle_{12345} \\ - |01000\rangle_{12345}-|01110\rangle_{12345}+\sqrt{2}|00000\rangle_{12345} \\ -\sqrt{2}|00110\rangle_{12345}+\sqrt{2}|01010\rangle_{12345}- \sqrt{2}|01100\rangle_{12345}\\ + |10011\rangle_{12345}+ |10101\rangle_{12345}+|11111\rangle_{12345} \\ + |11101\rangle_{12345}+|00011\rangle_{12345}+|00101\rangle_{12345}\\-|01011\rangle_{12345}-|01101\rangle_{12345} + \sqrt{2}|00011\rangle_{12345} \\ + \sqrt{2}|00101\rangle_{12345}+ \sqrt{2}|01001\rangle_{12345} -\sqrt{2}|01111\rangle_{12345}\big].
\end{multline}

\emph{\textbf{Step 4: Measurement of Qubit 4.}} 
Bob performs a measurement on qubit 4 on a computational basis. Depending on the outcome of the measurement (either $|0\rangle$ or $|1\rangle$), the state of the system collapses accordingly. 

If qubit 4 is measured to be in state $|0\rangle$, the state of the system becomes:

Substituting the expression for $|\omega\rangle_{12345}$ , the resulting state becomes: 

\begin{multline} 
|\omega\rangle_{12345} = \frac{1}{4} \big[|10000\rangle_{12345} + |11000\rangle_{12345} + |00000\rangle_{12345} \\ - |01000\rangle_{12345} \ + \sqrt{2}|00000\rangle_{12345} - \sqrt{2}|00110\rangle_{12345} \\+ |10011\rangle_{12345} + |11111\rangle_{12345} + |00011\rangle_{12345} \\ - |01011\rangle_{12345} + \sqrt{2}|00011\rangle_{12345} - \sqrt{2}|00101\rangle_{12345}\big]. 
\end{multline}

If qubit 4 is measured to be in state $|1\rangle$, the state of the system becomes:
\begin{multline}
|\omega\rangle_{12345} = \frac{1}{4} \big[|10110\rangle_{12345} + |11110\rangle_{12345} + |00110\rangle_{12345} \\ - |01110\rangle_{12345} \ + \sqrt{2}|00110\rangle_{12345} + \sqrt{2}|01010\rangle_{12345} \\ + |10101\rangle_{12345} + |11101\rangle_{12345} + |00101\rangle_{12345} \\ - |01101\rangle_{12345} + \sqrt{2}|00101\rangle_{12345} + \sqrt{2}|01001\rangle_{12345}\big]. 
\end{multline}

\emph{\textbf{Step 5: Applying CNOT Gates to Alice's and Bob's States.}} 
Alice's initial state is given by:
\begin{align}
    |\phi\rangle_A = \alpha |0\rangle_A + \beta |1\rangle_A,
\end{align}
and Bob's initial state is:
\begin{align}
    |\phi\rangle_B = \gamma |0\rangle_B + \eta |1\rangle_B.
\end{align}

The CNOT gate acts with Alice's qubit as the control and Bob's qubit as the target. The combined initial state of Alice and Bob is:
\begin{multline}
    |\psi\rangle_{AB} = |\phi\rangle_A \otimes |\phi\rangle_B \\ = (\alpha |0\rangle_A + \beta |1\rangle_A) \otimes (\gamma |0\rangle_B + \eta |1\rangle_B).
\end{multline}

Expanding this:
\begin{multline}
    |\psi\rangle_{AB} = \\ \alpha \gamma |00\rangle_{AB} + \alpha \eta |01\rangle_{AB} + \beta \gamma |10\rangle_{AB} + \beta \eta |11\rangle_{AB}.
\end{multline}

After applying the CNOT gate, the state is as follows:
\begin{multline}
    U_{\text{CNOT}} |\psi\rangle_{AB} = \\ \alpha \gamma |00\rangle_{AB} + \alpha \eta |01\rangle_{AB} + \beta \gamma |11\rangle_{AB} + \beta \eta |10\rangle_{AB}.
\end{multline}

The final state after the CNOT operation is therefore:
\begin{multline}
    |\psi'\rangle_{AB} = \\ \alpha \gamma |00\rangle_{AB} + \alpha \eta |01\rangle_{AB} + \beta \gamma |11\rangle_{AB} + \beta \eta |10\rangle_{AB}.
\end{multline}

\emph{\textbf{Step 6: Applying Hadamard Gates.}}
%
The state of the system before applying the Hadamard gates is:
\begin{multline}
    |\psi'\rangle_{AB} = \\ \alpha \gamma |00\rangle_{AB} + \alpha \eta |01\rangle_{AB} + \beta \gamma |11\rangle_{AB} + \beta \eta |10\rangle_{AB}.
\end{multline}

The Hadamard gate is applied to Alice's qubit, which transforms the basis states as follows:
\begin{align}
    H|0\rangle &= \frac{1}{\sqrt{2}} \left(|0\rangle + |1\rangle\right), \\
    H|1\rangle &= \frac{1}{\sqrt{2}} \left(|0\rangle - |1\rangle\right).
\end{align}

After applying the Hadamard gate to Alice's qubit, the state becomes:
\begin{multline}
    |\psi''\rangle_{AB} = H_A \otimes I_B \, |\psi'\rangle_{AB} \\
    = \frac{1}{\sqrt{2}} \Big[ \alpha \gamma \left(|0\rangle + |1\rangle\right) \otimes |0\rangle 
    + \alpha \eta \left(|0\rangle + |1\rangle\right) \otimes |1\rangle \\
    \quad + \beta \gamma \left(|0\rangle - |1\rangle\right) \otimes |1\rangle 
    + \beta \eta \left(|0\rangle - |1\rangle\right) \otimes |0\rangle \Big].
\end{multline}

Expanding this expression, we get:
\begin{multline}
    |\psi''\rangle_{AB} = \frac{1}{\sqrt{2}} \Big[
    \alpha \gamma |00\rangle + \alpha \eta |01\rangle 
    + \alpha \gamma |10\rangle + \alpha \eta |11\rangle \\
    + \beta \eta |00\rangle - \beta \gamma |01\rangle 
    - \beta \eta |10\rangle + \beta \gamma |11\rangle
    \Big].
\end{multline}

Simplifying, the state becomes:
\begin{multline}
    |\psi''\rangle_{AB} = \frac{1}{\sqrt{2}} \Big[
    (\alpha \gamma + \beta \eta)|00\rangle + (\alpha \eta - \beta \gamma)|01\rangle \\
    + (\alpha \gamma - \beta \eta)|10\rangle + (\alpha \eta + \beta \gamma)|11\rangle
    \Big].
\end{multline}

\emph{\textbf{Step 7: Measurement and recovery.}} 
Alice and Bob perform measurements on their respective qubits and communicate the results via a classical channel. Depending on the outcomes, unitary corrections (Pauli gates X and Z) are applied to recover the original states $|\phi\rangle_A$ and $|\phi\rangle_B$ at the target locations.
\subsection{Bidirectional Quantum Teleportation using a Cluster-Bell State:}
The second quantum circuit used in our multihop bidirectional quantum teleportation study is presented in this section. This circuit compares its performance with other quantum channels by using a linear cluster state as an entangled resource. The goal is to find the most effective quantum channel to find the shortest path in bidirectional communication.

For an $n$-qubit system, a linear cluster state, a significant type of multipartite entangled state, can be represented as follows:
\begin{align}
|C_n\rangle = \frac{1}{2^\frac{n}{2}}\bigotimes_{d = 1}^{n} (|0\rangle_d \sigma^{d +1}_z + |1\rangle_d)
\end{align}
where the computational basis states of the d-th qubit are represented by $|0\rangle_d $ and $|1\rangle_d$, and the Pauli Z operator acting on the $(d+1)-th$ qubit is represented by $\sigma^{d +1}_z $.
The fundamental multipartite entanglement structure of this state and good noise tolerance make it a strong contender for quantum communication. Our goal is to examine the accuracy and effectiveness of quantum information transport across multihop by building the communication network using linear cluster states. By using this method, we may investigate whether the cluster Bell state is more stable to use to optimize the path than other quantum entangled states like the GHZ-Bell states used before.
A four-qubit cluster state used as a quantum channel between Alice and Bob is in the state: 
\begin{multline}
    |C\rangle_{123456} = \frac{1}{2} \left( |0000\rangle + |0011\rangle + |1100\rangle_{1234} + |1111\rangle_{1234}\right)\\\otimes \frac{1}{\sqrt{2}} \left( |00\rangle + |11\rangle \right)_{56}
\end{multline}
\begin{multline}    
    |C\rangle_{123456} = \frac{1}{2\sqrt{2}} \big( |000000\rangle + |001100\rangle + |110000\rangle + |111100\rangle \\ + |000011\rangle + |001111\rangle + |110011\rangle + |111111\rangle
    \big)_{123456}.
\end{multline}

\begin{figure*}[!t]
    \centering
    \includegraphics[width=1\textwidth]{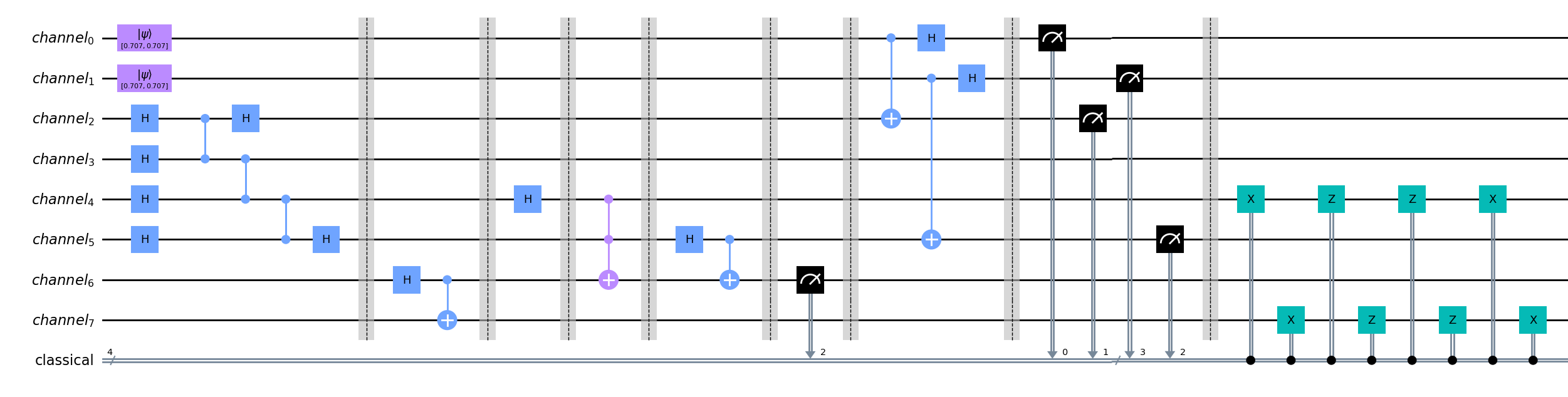}
    \caption{Quantum circuit for bidirectional quantum teleportation using a Cluster-Bell state as the shared entangled resource.}
    \label{fig3}
\end{figure*}

Following the method described in Appendix (\ref{12}), we analyze the quantum states corresponding to Alice and Bob, which are represented as $|\phi\rangle_A$ and $|\phi\rangle_B$, respectively. These states will be expressed in the quantum circuit shown in Fig.\ref{fig2} and will serve as the basis for the operations that follow. The circuit in Fig.\ref{fig3} provides a detailed visual representation of the procedure by encoding the interactions and transformations necessary for the protocol.

The final states of the system are obtained by applying the sequence of quantum operations. A Hadamard gate on qubit 4 initiates the procedure by superposing its base states. A controlled-controlled-NOT (CCNOT) gate then acts on qubits 3, 4, and 5 to create entanglement between them. Qubit 4 then passes through another Hadamard gate, which further modifies its state in relation to previous interactions. Finally, correlations between the states of qubits 4 and 5 are introduced by applying a CNOT gate between them. The system's final states are obtained by these operations and the required measurements.
The measurement results reveal a distribution of outcomes across multiple quantum states, indicating the richness of the underlying quantum system. The observed results include \(1111\), \(0100\), \(1110\), \(0010\), \(0000\), \(1101\), \(0001\), \(0011\), \(1011\), \(1100\), \(0110\), \(1000\), \(1001\), \(0101\), and \(1010\). These results show how our experimental setup allows for an extensive number of measurement settings. Because quantum mechanics is fundamentally probabilistic, each result relates to a particular quantum state or set of states. This varied distribution offers important insights for additional research by highlighting consistency and variability within the quantum system under study.

\section{Quantum Dijkstra algorithm and Quantum walk for Optimal Path in Multihop BQT}  
In this section, we introduce BQT-MDQW, a quantum approach that leverages quantum walks \cite{2003} and the quantum Dijkstra's Algorithm \cite{2022} to find the best path in multihop bidirectional quantum teleportation (BQT) networks. Qiskit and Python are used to build a system that combines the concepts of quantum computing with graph optimization techniques. The suggested method finds the shortest path between the source and destination nodes in large-scale quantum networks as was done in the article \cite{2024, 2020} if they combined phase estimation with a modified Grover search algorithm.
The quantum walk protocol improves reliability in noisy quantum environments by introducing a reliable network access technique and including the prevention of errors. The method creates quantum communication connections between nodes using entangled channels, which are represented as cluster bell states, the Werner-bell state, and the GHZ-bell state. This ensures coherence during the teleportation process. In networks of up to $200$ nodes, this hybrid framework combining Dijkstra's algorithm and quantum walks offers an effective way to solve scaling issues while optimizing path finding.
Throughput calculations and visualization applications for network structure analysis are included in the implementation to assess performance across different network sizes. An unusual approach to solving challenging optimization issues in quantum communication networks is offered by combining quantum walks and graph theory.

In order to execute the quantum Dijkstra's algorithm, the input graph is represented by a quantum circuit, and the required computations are carried out using quantum gates. In addition, to improve the efficiency of the quantum circuit and obtain the best possible answer, traditional approaches including optimization techniques and graph-theory algorithms are used. By comparing it with the classical Dijkstra's algorithm, the quantum Dijkstra's algorithm's performance is evaluated, with particular focus paid to measures such as time complexity and the number of quantum gates used. The quantum walk circuit implemented in the code uses entangled quantum channels and controlled quantum gates to simulate traversal on a network graph. The input graph is encoded with nodes represented as quantum registers and edges as quantum channels prepared using a composite entangled W-Bell state, GHZ-Bell state and Cluster-Bell state.
The quantum walk is performed along a predefined path, where for each step, the quantum channel for the corresponding edge is incorporated into the circuit and a \(CX(q_i, q_j)\) operation is applied between the quantum registers of the current node \(i\) and its neighbor \(j\). After completion of the walk for the specified number of steps, measurements \(M(q_i)\) are made on all qubits, and error mitigation is implemented through conditional recovery operations. This circuit integrates entangled quantum channels with the classical Dijkstra algorithm for path-finding, offering a scalable and adaptable approach to simulate quantum walks in large networks while leveraging the coherence of quantum entanglement for efficient information transfer.

\section*{Algorithm: BQT-MDQW}
\begin{enumerate}
    \item[Input:] Graph $G$ with nodes $V$ and edges $E$, source $s$, target $t$
    \item[Output:] Forward and backward quantum paths

    \item \textbf{Setup}
    \begin{enumerate}
        \item Import required libraries (e.g., Qiskit, NetworkX, Matplotlib, NumPy)
        \item Initialize quantum register $qr$ and classical register $cr$
    \end{enumerate}

    \item \textbf{Entanglement Preparation}
    \begin{enumerate}
        \item Set Alice's qubit: $q_A = \alpha|0\rangle + \beta|1\rangle$
        \item Set Bob's qubit: $q_B = \gamma|0\rangle + \delta|1\rangle$
        \item Prepare W-state:
        \begin{itemize}
            \item Apply $R_y(\pi/2)$ to $q_A$
            \item Apply Controlled-Hadamard from $q_A$ to $q_B$
            \item Apply CNOT from $q_B$ to auxiliary qubit
            \item Apply X gate to auxiliary qubit
        \end{itemize}
        \item Prepare Bell-state:
        \begin{itemize}
            \item Apply Hadamard to $q_A$
            \item Apply CNOT from $q_A$ to $q_B$
        \end{itemize}
    \end{enumerate}

    \item \textbf{Quantum Walk}
    \begin{enumerate}
        \item For each step up to max\_steps:
        \begin{itemize}
            \item For each edge $(u,v)$ in path:
            \item[] Apply CNOT from $qr[u]$ to $qr[v]$
        \end{itemize}
        \item Measure $qr$ into $cr$
    \end{enumerate}

    \item \textbf{Path Finding}
    \begin{enumerate}
        \item Compute forward path $P_{forward}$ using Quantum Dijkstra ($s$ to $t$)
        \item Compute backward path $P_{backward}$ using Quantum Dijkstra ($t$ to $s$)
    \end{enumerate}

    \item \textbf{Visualization}
    \begin{enumerate}
        \item Draw graph $G$ using NetworkX
        \item Highlight $P_{forward}$ in red, $P_{backward}$ in green
    \end{enumerate}

    \item \textbf{Execution}
    \begin{enumerate}
        \item Construct graph $G$
        \item Prepare entanglement (step 2)
        \item Find paths (step 4)
        \item Perform quantum walks (step 3)
        \item Visualize results (step 5)
    \end{enumerate}
\end{enumerate}
In our work, we used \textbf{BQT-MDQW}, which combines the quantum walk and the quantum Dijkstra algorithm. This method is an important part of the bidirectional quantum teleportation (BQT) process. It helps simulate how quantum states move across a network, which we represent as a graph. By creating entangled connections between nodes using W-Bell states, the quantum walk uses the principles of superposition and entanglement to transfer information more efficiently.
Each step of the walk involves a series of entanglement-based operations and CNOT gates that replicate the flow of information along a predefined path. Notably, the quantum walk operates on a graph with nodes labeled from Alice to Bob, utilizing the shortest path between them to optimize resource use and fidelity. This simulation emphasizes that quantum information transfer is not a straightforward process; instead, it involves intermediate quantum states dynamically interacting across a network. Using W-states in the channel preparation ensures robustness against decoherence, making the quantum walk a crucial component in simulating effective quantum network communication.

\section{Performance Evaluation}
\begin{figure*}[!t]
        \includegraphics[width=.3\linewidth]{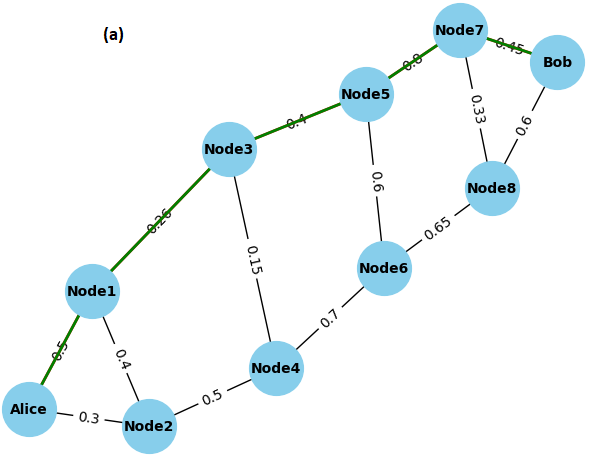}\hspace{0.5cm}
        \includegraphics[width=.3\linewidth]{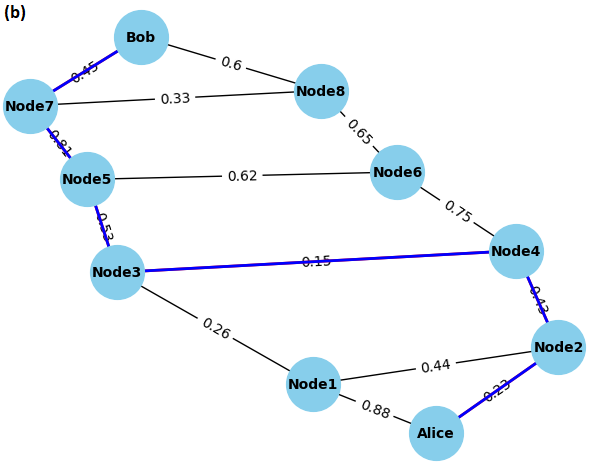}\hspace{0.5cm}
        \includegraphics[width=.3\linewidth]{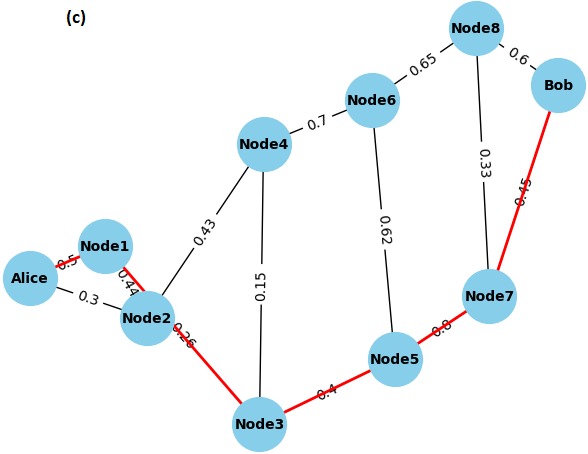}
    \caption{Shortest path identification using BQT-MDQW three different entangled states: ($a$) W-Bell state, ($b$) GHZ-Bell state, and ($c$) Cluster-Bell state. Each panel shows the identified shortest path and the corresponding quantum walk probability distribution.}
    \label{fig4}
\end{figure*}

In this implementation, we randomly initialized the edge weights in the network graph to create paths between the nodes, simulating a realistic and dynamic quantum network. The graph consists of $10$ nodes, including Alice and Bob as endpoints, connected by intermediary nodes that form a network of quantum channels. These weights, representing the resource cost or potential communication errors associated with each channel, influence the construction of paths between nodes and serve as input for the quantum Dijkstra algorithm. The primary role of the quantum walk in this context is to simulate the traversal of quantum information across the network along the paths identified by the quantum Dijkstra algorithm. The quantum walk facilitates the propagation of quantum states and serves as a tool for studying the dynamics of quantum teleportation in networked systems. The code introduces the "{$find$\_$shortest$\_$path$}" function, which employs quantum Dijkstra's algorithm via the NetworkX library to determine the shortest path. This function takes a graph as in Fig.\ref{fig4}, a starting node, and an ending node as inputs, computing the shortest path between these two nodes according to the edge weights.


The BQT-MDQW code ultimately identifies the shortest path between two locations by analyzing the measurement results. It employs Quantum Walk to perform this calculation. Each step of the walk consists of a series of entanglement-based operations that help determine which results align with a route from the starting point to the destination. 

The results of this study show that quantum computing has a lot of promise for speeding up the weighted graph shortest path search. Inspired by quantum Dijkstra's algorithm, a quantum algorithm for the shortest path issue was implemented. The solution has been successfully tested on real quantum devices and simulators using the IBM Qiskit platform. The quantum method may exceed the classical Dijkstra's algorithm by an exponential amount, according to simulation results, and studies on actual quantum devices verify that it is possible to implement it on short-term quantum hardware. By analyzing the measurement results, the code carefully finds the shortest path between two places. It uses the Quantum Walk approach to make this computation easier. To determine which outcomes correspond to a feasible path from the starting point to the destination, each iteration of the walk consists of a sequence of entanglement-based methods.

\begin{figure}[htp]
    \centering
    \includegraphics[width=8cm]{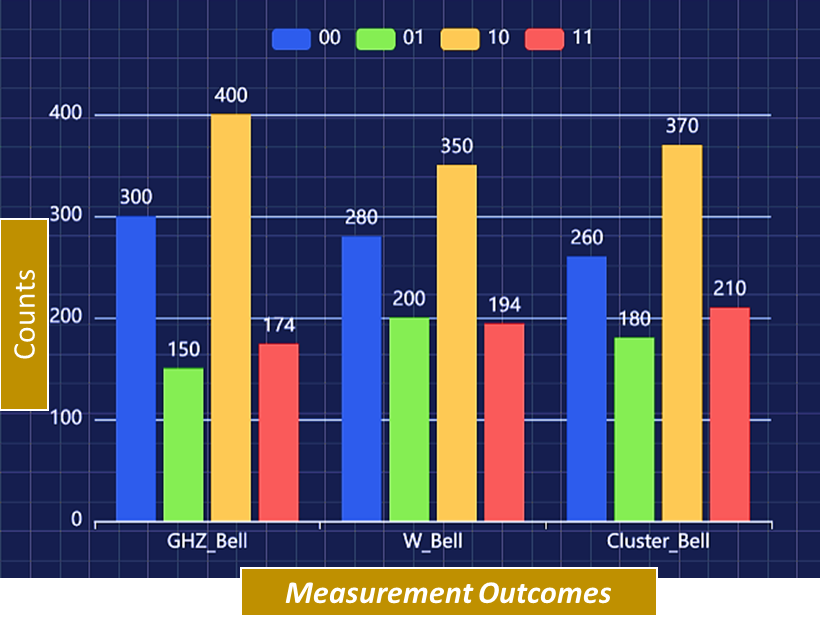}
    \caption{Comparison of Quantum Walk Results for Different Entangled States in Bidirectional Quantum Teleportation}
    \label{figwalk}
\end{figure}

For BQT between Alice and Bob, the figure (\ref{figwalk}) shows the MCDM shortest pathways and some of the quantum walk outcomes using three distinct entangled states: the Cluster-Bell state (bottom), the W-Bell state (middle), and the GHZ-Bell state (top). The best route for quantum communication is shown in each part, along with the appropriate quantum walk outcomes in both directions. To illustrate how different entanglement architectures affect teleportation efficiency, the color-coded zones draw attention to differences in the quantum state distributions. The GHZ-Bell and Cluster-Bell states show different patterns of quantum state propagation, while the W-Bell state has a more balanced distribution.

This study shows how quantum computing could transform graph algorithms, particularly for complexity and optimization problems. Despite its potential, there are still many obstacles to overcome, such as improving scalability, decreasing error rates, and optimizing quantum circuits. However, compared to the classical technique, the quantum Dijkstra's algorithm's time complexity was significantly reduced when implemented with the Qiskit library. The quantum version of Dijkstra's algorithm achieves $O(\log (N)^2)$, where N is the number of nodes in the graph, while the classical version has a time complexity of $O(N^2)$. These findings show the potential of quantum computing as a quicker and more effective method of resolving the shortest path issue.

\subsection{Simulation Methodology} 
We perform our simulation based on an off-the-shelf quantum network simulation framework called SimQN \cite{2023}. The software and hardware configuration of the simulation platform is AMD Ryzen $7$ $3700$X $3.6$GHz CPU, $32$GB RAM, and OS Windows $10$ $64$bits. We follow \cite{huang2022socially} to generate a quantum network using the Waxman model \cite{1998routing}. We randomly deploy $200$ quantum nodes in a rectangle area of $2000$ km by $4000$ km and randomly assign them to $10$ S-D pairs in the network. We use the Waxman model to build an edge with probability $\delta e^{-l\left(u,v\right)/\varepsilon L}$, where $L$ is the largest distance between two nodes, $\delta=0.90$ is the probability of creating an edge between two nodes, $\varepsilon=0.01$ is a parameter controlling the sensitivity of edge creation to distance and $l(u,v)$ is the distance between node $u$ and node $v$. It ensures that the average distance between two adjacent nodes is about 100 km, typical for fibers \cite{yin2012quantum}. We conducted $1000$ runs for each set of parameters, calculated the average results, and set the qubit sending rate at $1000$ qubits/second across the network. To verify network connectivity, we set a classical delay of $0.05$s. 
Each node is equipped with $50$ units of quantum memory, and the qubits had a drop rate of $0.03$. We defined the probability of swapping at the adjacent node to $0.98$, ensuring reliable performance throughout the system.

We analyze the performance of unidirectional (Alice-to-Bob) and bidirectional (Alice-to-Bob and Bob-to-Alice) teleportation using key metrics such as throughput, end-to-end fidelity, and memory utilization. Throughput refers to the expected number of successfully generated end-to-end qubits, focusing on the protocol's connectivity capability without factoring in the fidelity of individual links. On the other hand, end-to-end fidelity evaluates the quality of the entangled connections established across the network, ensuring reliable quantum state transfer. Memory utilization is crucial in assessing unidirectional and bidirectional teleportation efficiency, as it quantifies the ratio of entangled pairs used to the total pairs available. Higher memory utilization indicates more effective resource usage, directly reflecting the performance and scalability of the teleportation scheme in quantum networks. We evaluated the performance of unidirectional and bidirectional quantum teleportation in different entangled states, including Werner states \cite{lee2000entanglement}, GHZ states, and Cluster-Bell states. 

\begin{figure*}
    \centering
    \subfigure[Throughput]{
        \begin{minipage}[t]{0.31\linewidth}
            \centering
            \includegraphics[width=1\textwidth]{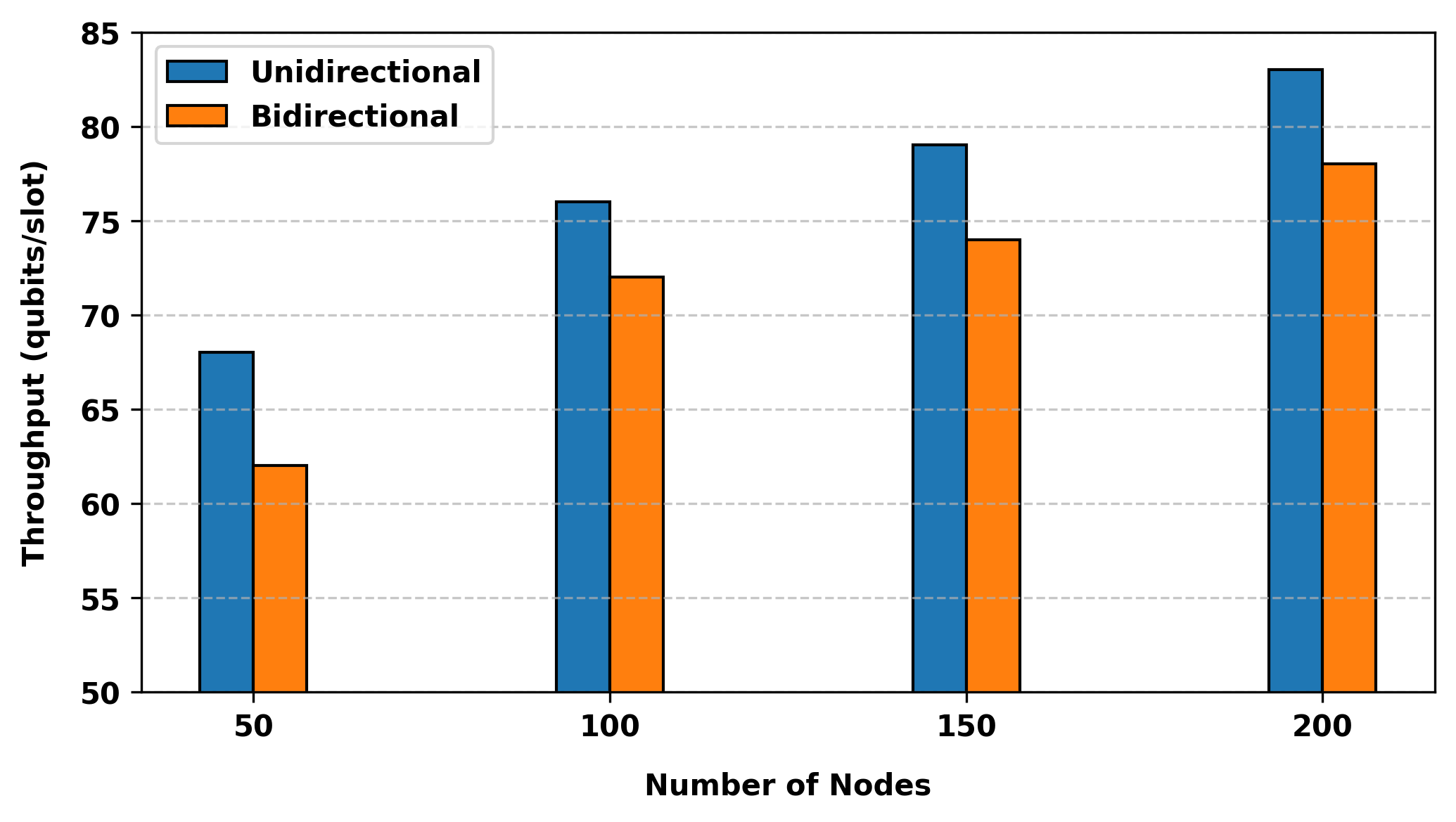}
        \end{minipage}
    } 
    \subfigure[Fidelity]{
        \begin{minipage}[t]{0.31\linewidth}
            \centering
            \includegraphics[width=1\textwidth]{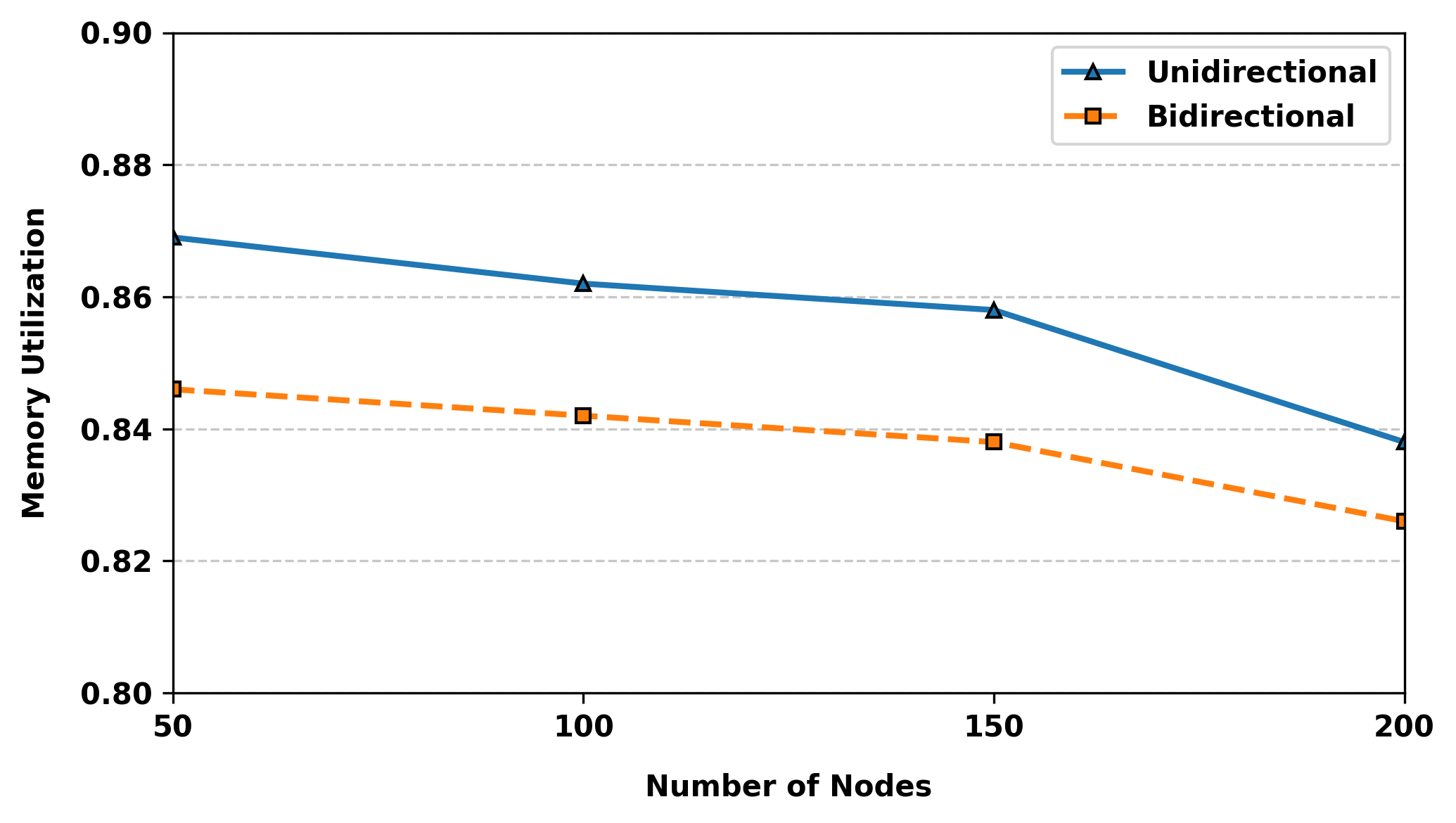}
        \end{minipage}
    } 
    \subfigure[Memory utilization]{
        \begin{minipage}[t]{0.31\linewidth}
            \centering
            \includegraphics[width=1\textwidth]{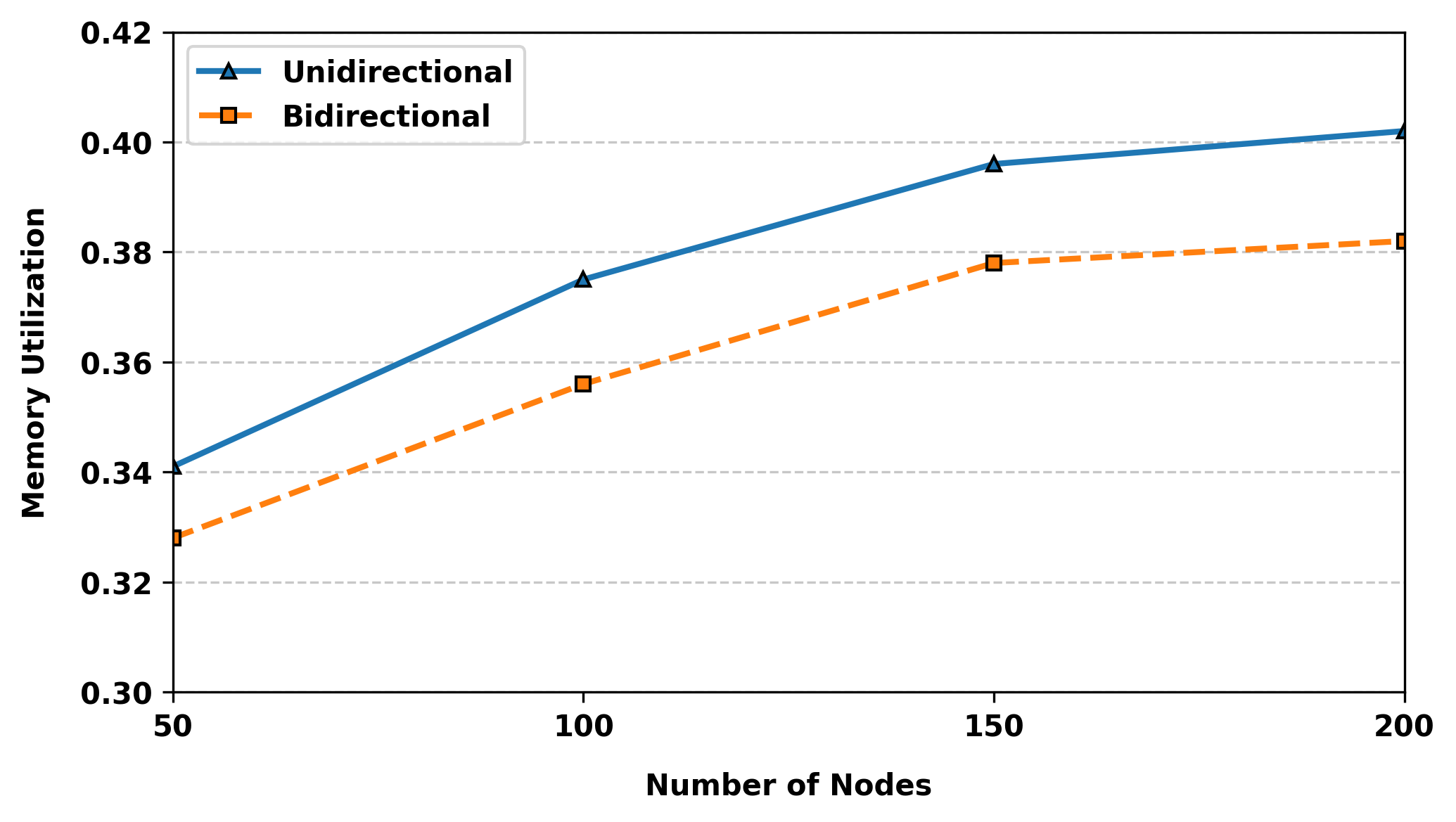}
        \end{minipage}
    }
    
    \vspace{5pt} 

    \subfigure[Throughput]{
        \begin{minipage}[t]{0.31\linewidth}
            \centering
            \includegraphics[width=1\textwidth]{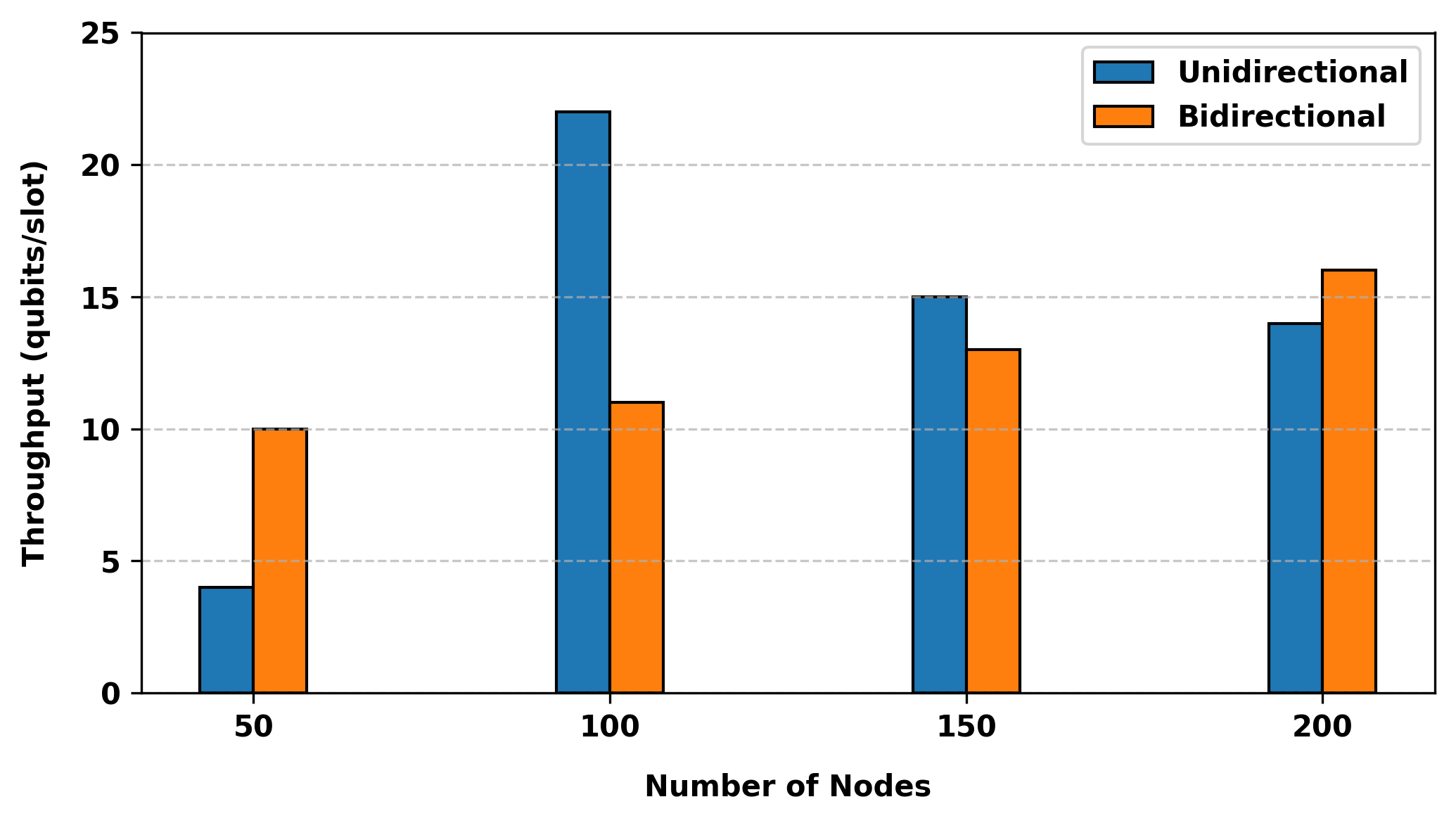}
        \end{minipage}
    } 
    \subfigure[Fidelity]{
        \begin{minipage}[t]{0.31\linewidth}
            \centering
            \includegraphics[width=1\textwidth]{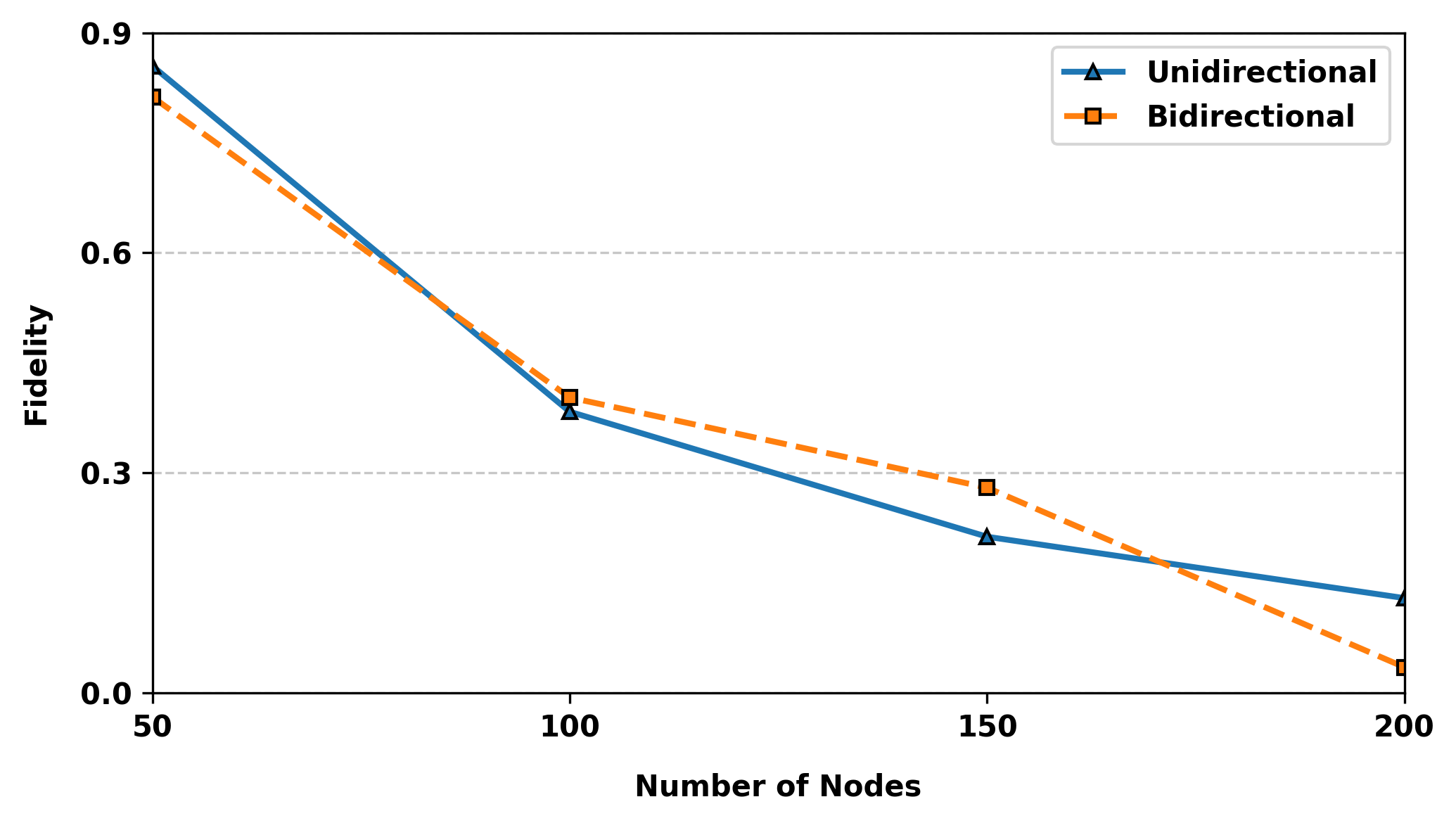}
        \end{minipage}
    } 
    \subfigure[Memory utilization]{
        \begin{minipage}[t]{0.31\linewidth}
            \centering
            \includegraphics[width=1\textwidth]{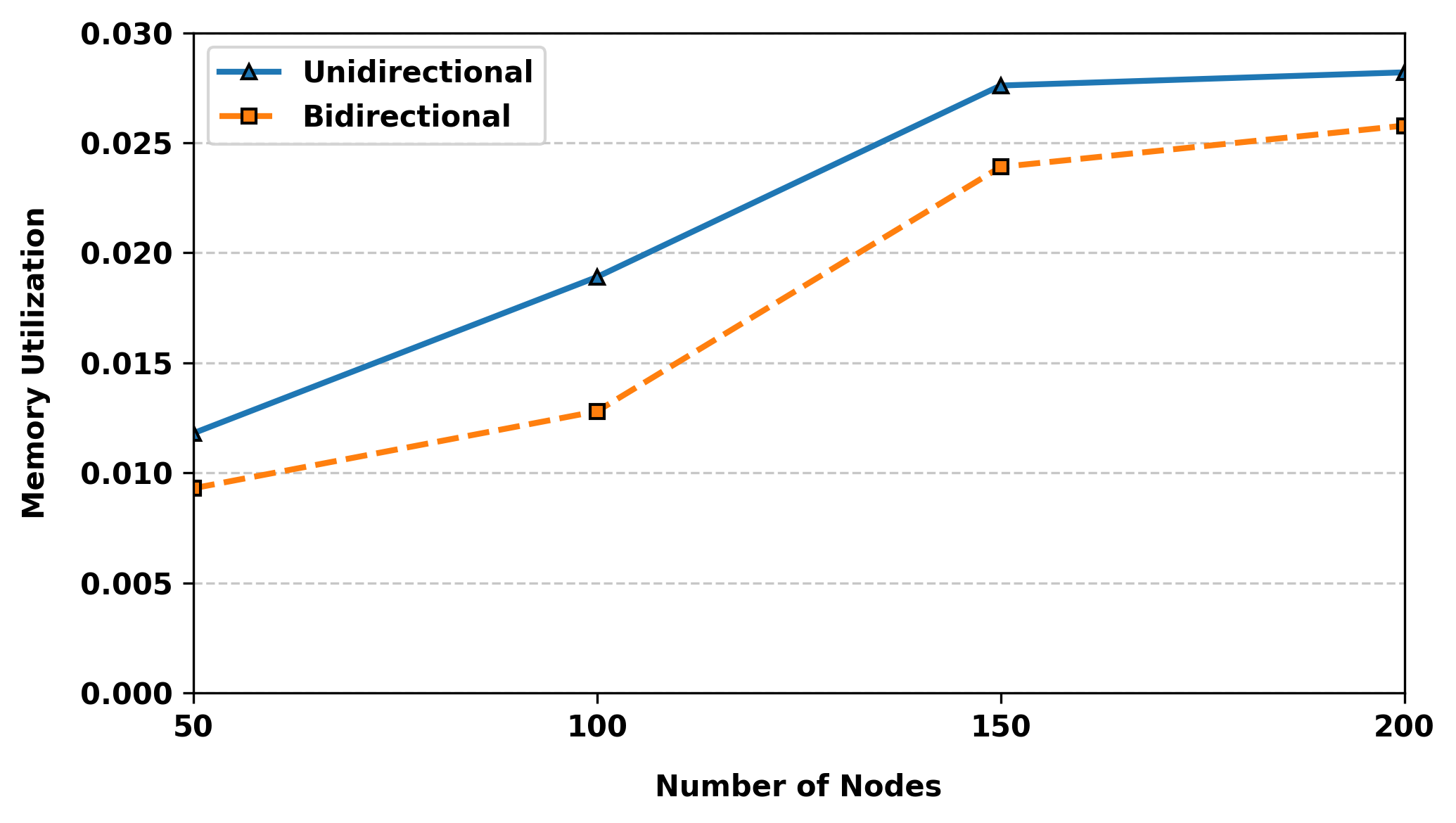}
        \end{minipage}
    }

    \caption{Simulation-based performance of W-Bell states concerning the number of nodes, illustrated in (a), (b), and (c). Simulation of BQT-MDQW with optimal routing and a dynamically created entangled W-Bell channel is shown in (d), (e), and (f).}
    \label{fig5}
\end{figure*}

\begin{figure*}
\subfigure[Throughput]{
\centering
\begin{minipage}[t]{0.32\linewidth}
\centering
\includegraphics[width=1\textwidth]{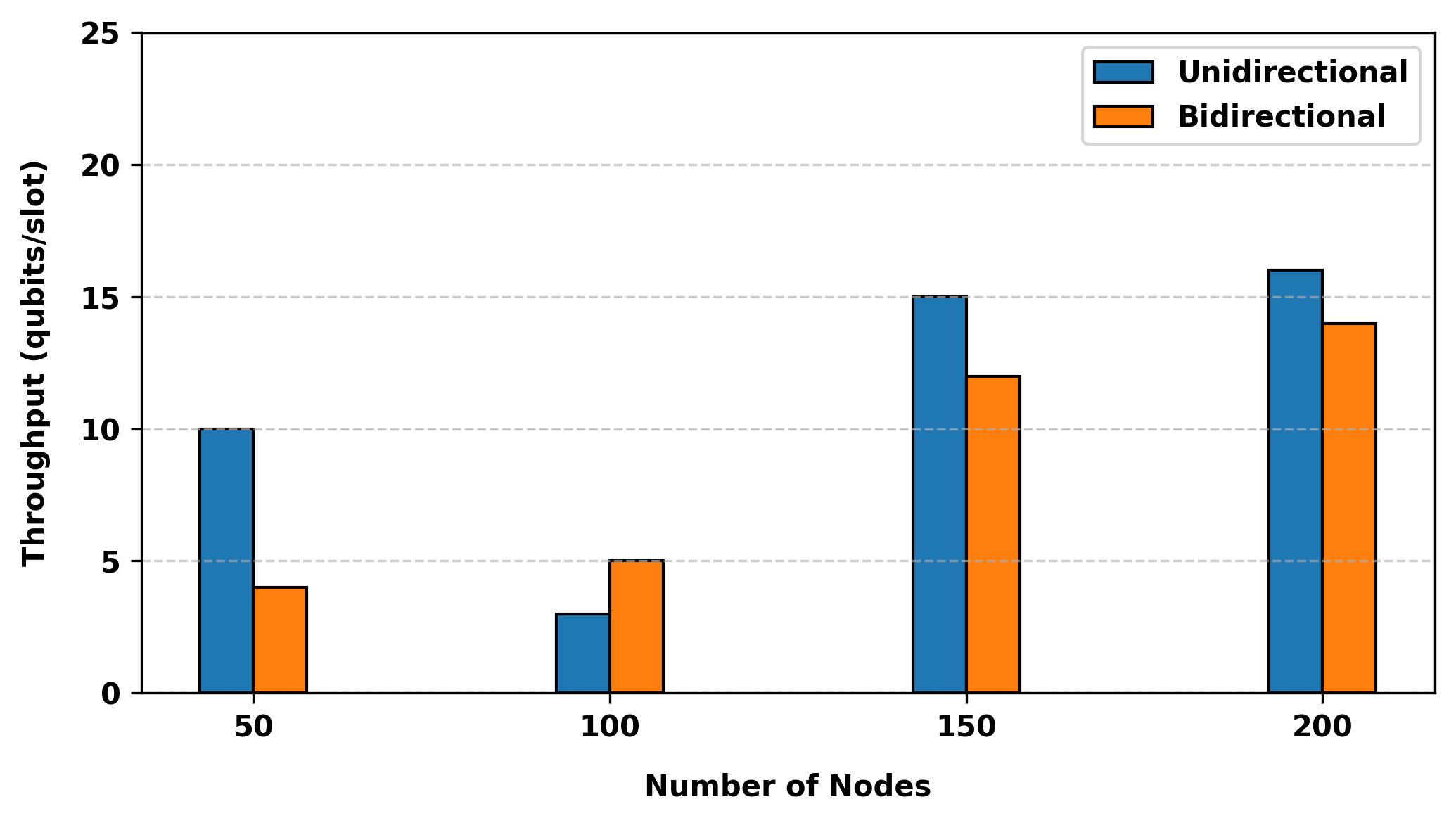}
\end{minipage}} 
\subfigure[Fidelity]{
\centering
\begin{minipage}[t]{0.32\linewidth}
\centering
\includegraphics[width=1\textwidth]{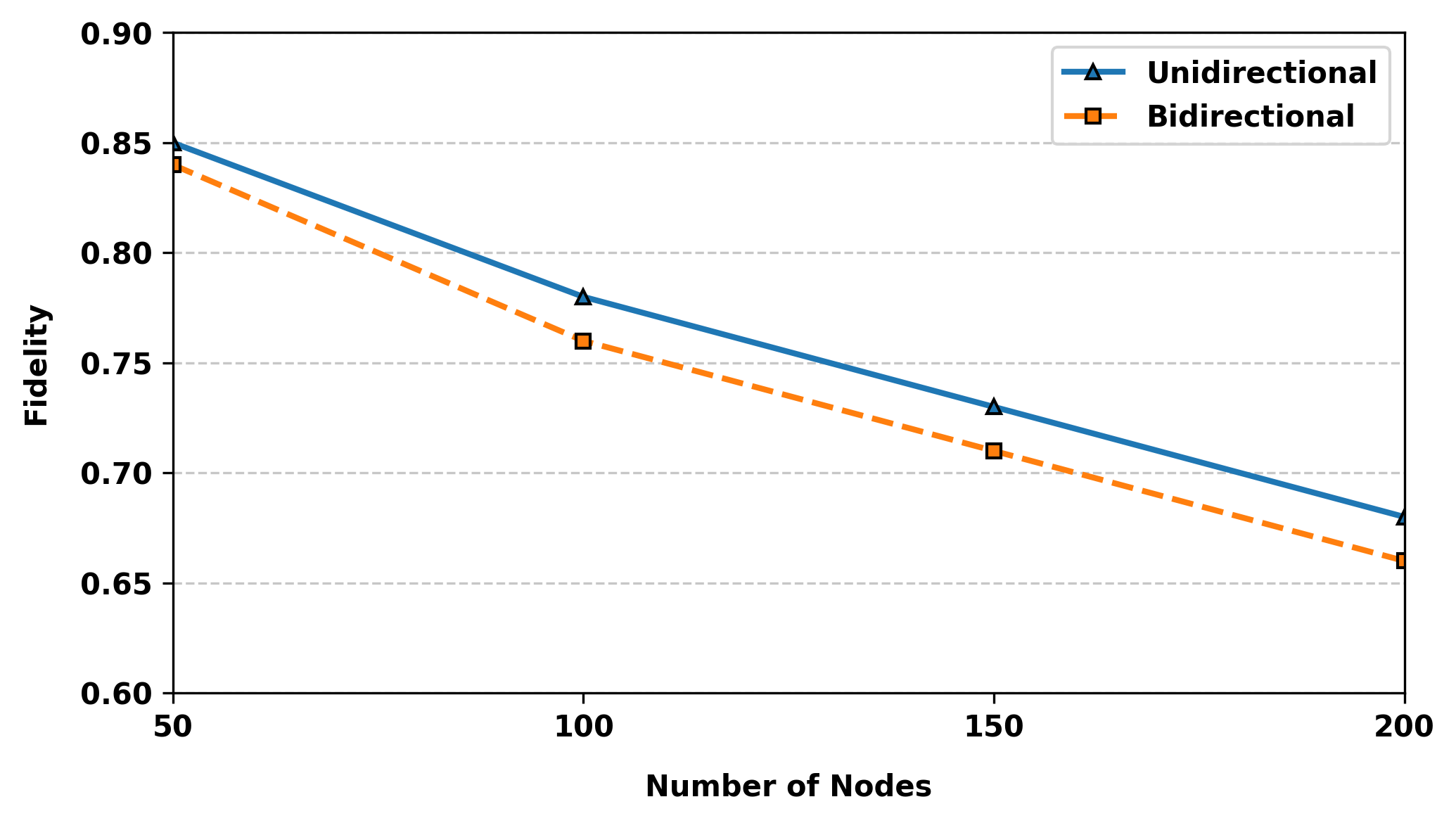}
\end{minipage}} 
\subfigure[Memory utilization]{
\begin{minipage}[t]{0.32\linewidth}
\centering
\includegraphics[width=1\textwidth]{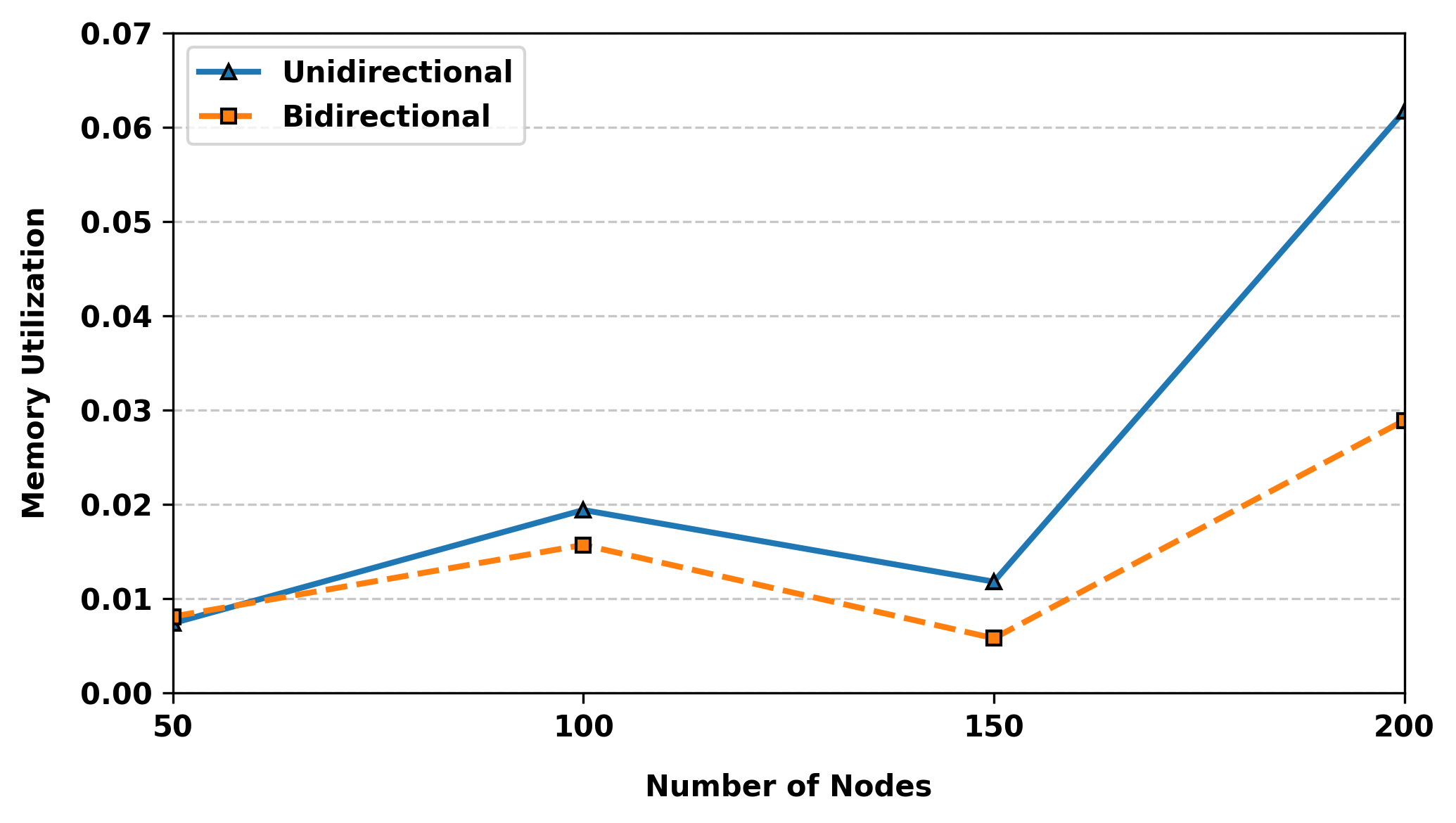}
\end{minipage}}%
\caption{BQT-MDQW simulation using dynamically generated entangled GHZ-Bell channel and optimal routing.}
\label{fig7}
\end{figure*}

\begin{figure*}
\subfigure[Throughput]{
\centering
\begin{minipage}[t]{0.32\linewidth}
\centering
\includegraphics[width=1\textwidth]{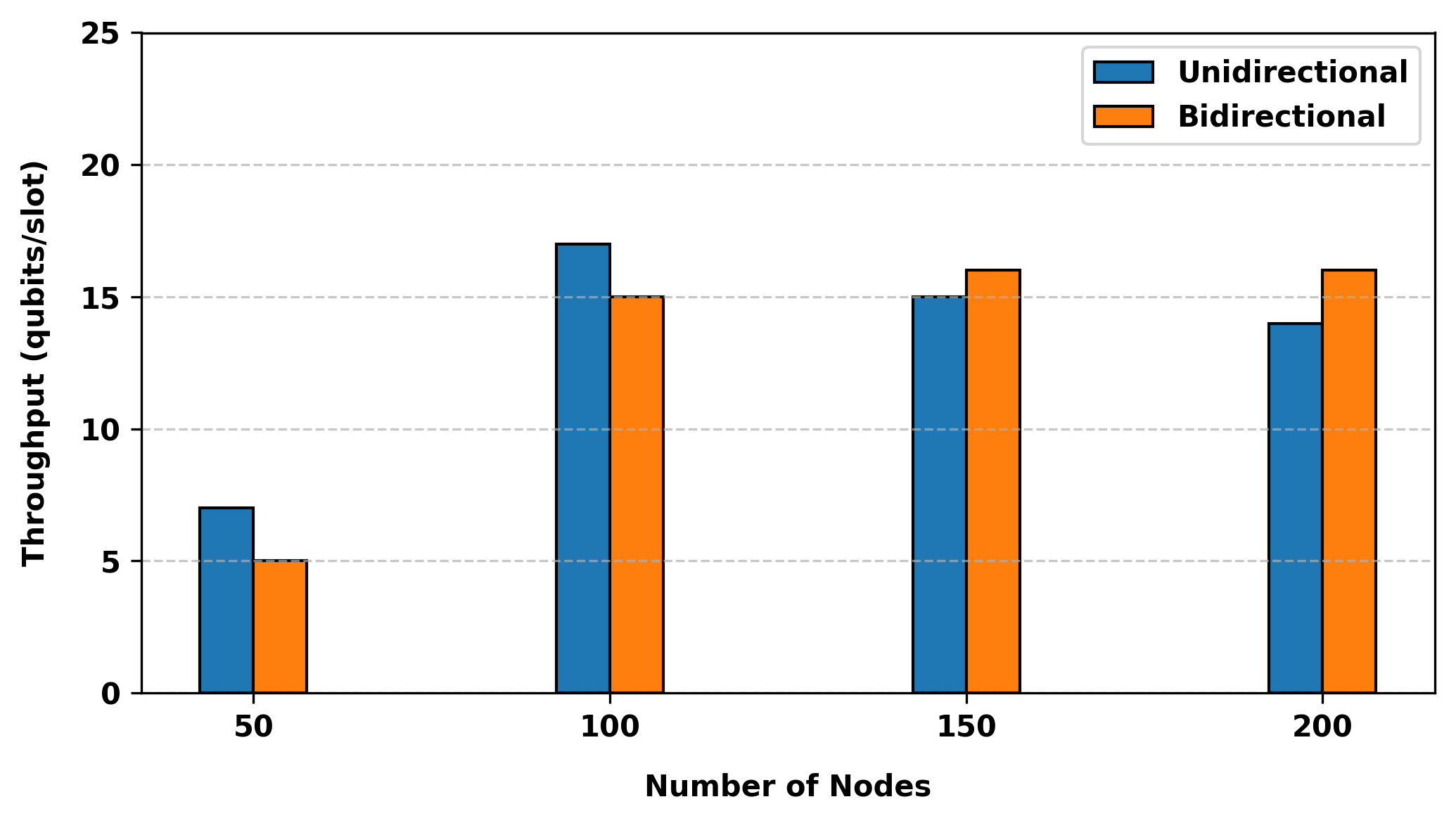}
\end{minipage}} 
\subfigure[Fidelity]{
\centering
\begin{minipage}[t]{0.32\linewidth}
\centering
\includegraphics[width=1\textwidth]{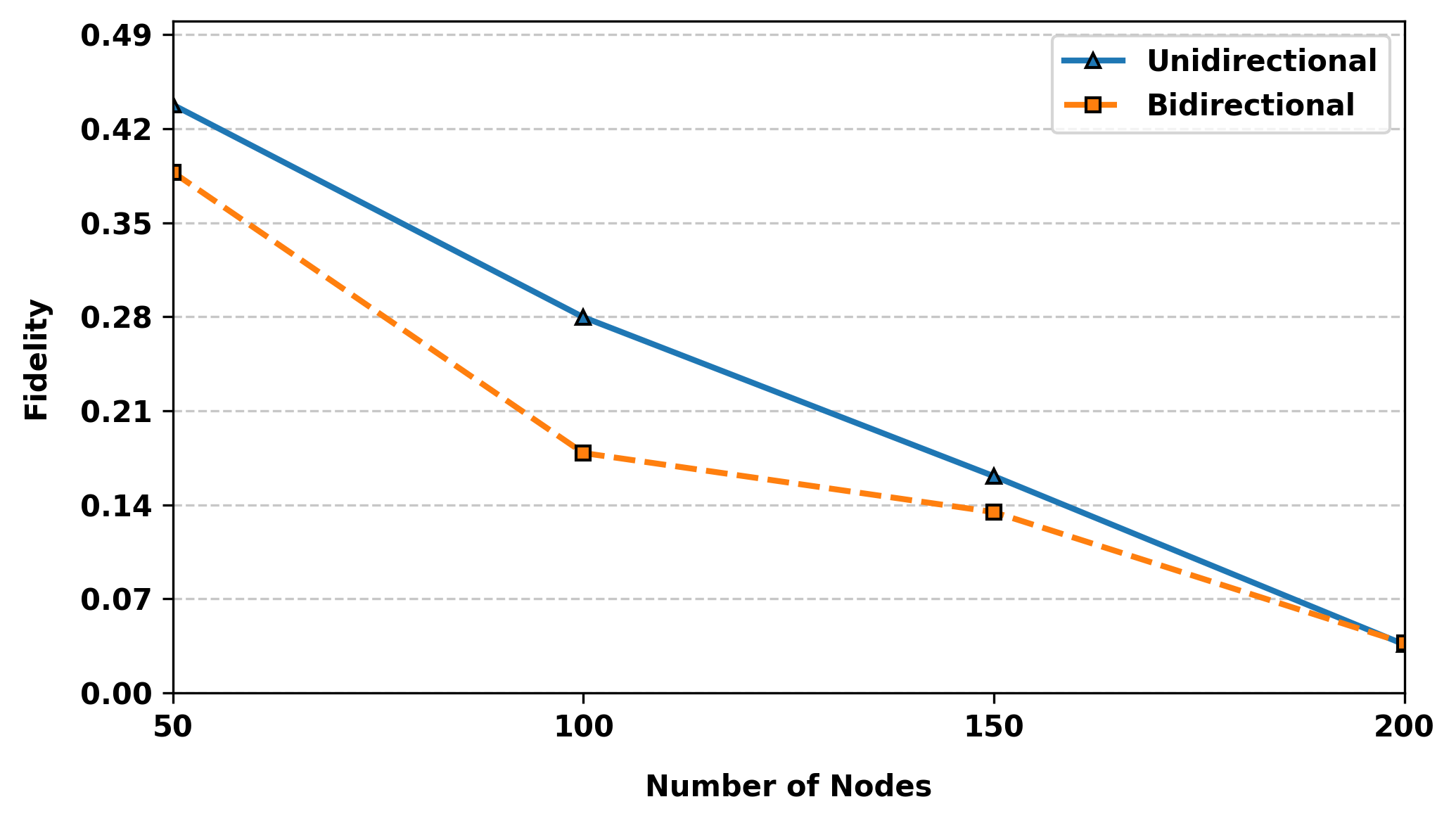}
\end{minipage}} 
\subfigure[Memory utilization]{
\begin{minipage}[t]{0.32\linewidth}
\centering
\includegraphics[width=1\textwidth]{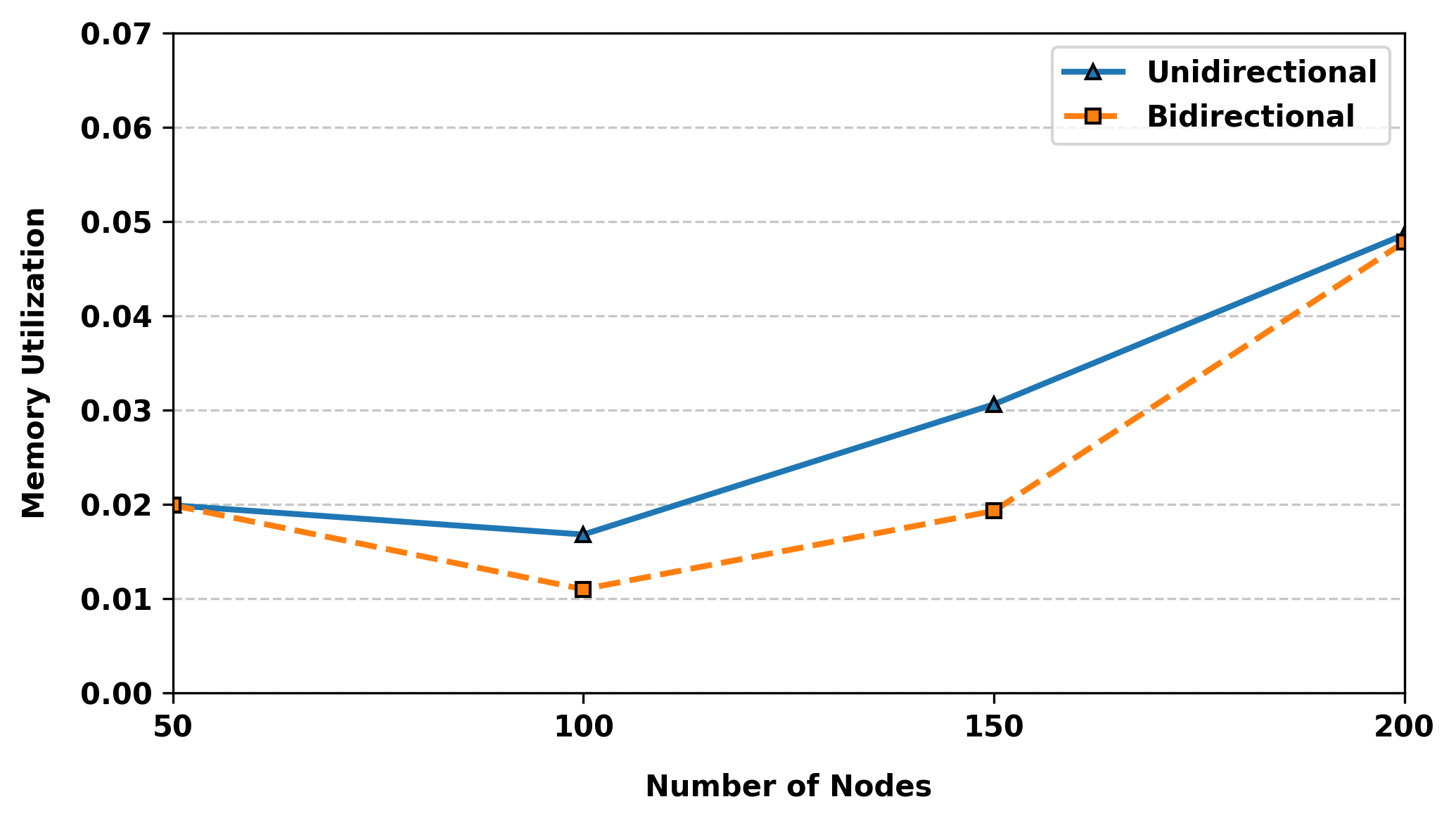}
\end{minipage}}%
\caption{Simulation of BQT-MDQW with optimal routing and a dynamically created entangled Cluster-Bell state}
\label{fig8}
\end{figure*}

\subsubsection{Simulation performance analysis of entangled states:} 
{\bf W-Bell state:} Figure \ref{fig5}(a) illustrates the performance of unidirectional and BQT-MDQW throughput, highlighting the advantages of unidirectional transmission in quantum networks. This is because unidirectional teleportation allows for dynamic path selection, optimizing routing decisions based on the availability of high-fidelity entanglement links, whereas bidirectional teleportation constrains routing choices by requiring paths to be viable in both directions simultaneously. Furthermore, errors in bidirectional teleportation propagate across both directions, increasing the need for complex error correction, while unidirectional teleportation minimizes decoherence risks by focusing on a single transmission at a time. By reducing contention, minimizing delays, and maximizing entanglement efficiency, unidirectional teleportation significantly higher throughput in quantum networks. Figure \ref{fig5}(b) demonstrates that unidirectional teleportation achieves higher fidelity than BQT-MDQW due to its more efficient use of entanglement resources. In bidirectional teleportation, the process depends on two high-fidelity entangled links simultaneously. If even one entangled pair has reduced fidelity, it can significantly degrade the success of the entire transmission. However, unidirectional teleportation avoids this constraint by selecting the highest-quality available link, ensuring a more reliable and higher-fidelity quantum state transfer. In Figure \ref{fig5}(c), unidirectional teleportation achieves higher quantum memory utilization than BQT-MDQW, particularly as the number of nodes increases. This is due to bidirectional teleportation that requires two simultaneous entangled links, which doubles memory usage and prolongs retention time while waiting for a viable two-way connection. This leads to memory bottlenecks and a higher risk of decoherence. Conversely, unidirectional teleportation immediately employs the highest-fidelity available link, reducing idle memory time and allowing for continuous transmission.\\ 

We study the performance of BQT-MDQW using W-Bell states as the entanglement channel and analyze the quantum network's throughput based on the number of qubits transmitted per time slot as in Figure \ref{fig5}(d). We simulate a dynamically generated network with up to \textbf{200} nodes in which Alice and Bob communicate using the shortest entangled path found using Dijkstra's algorithm. The teleportation protocol uses a composite W-Bell state to ensure robust entanglement distribution and error mitigation. The simulation results, displayed in the throughput vs. network size graph, demonstrate clear performance differences between unidirectional and bidirectional quantum teleportation. BQT shows increased efficiency as the network size grows, while unidirectional teleportation shows higher throughput for smaller network sizes. This work opens the door for optimized quantum communication protocols by showcasing the potential of W-Bell states for scalable quantum networks.

An important parameter to measure the accuracy of entanglement-based communication methods is the fidelity of quantum teleportation. In this work, we use a composite W-Bell entangled state as the quantum channel and examine the performance of a BQT protocol. The degree to which the transmitted quantum state resembles the state that is wanted is measured by the fidelity, which is denoted by \[F = \left| \langle \psi_{\text{ideal}} | \rho_{\text{actual}} | \psi_{\text{ideal}} \rangle \right|.\] 

We examine the fidelity of unidirectional and bidirectional teleportation techniques under quantum walk-assisted routing in a simulation on a 200-node network. The results, which are shown in Figure \ref{fig5}(e), show that BQT is more resistant against decoherence and retains higher fidelity as network sizes increase. The potential of W-Bell states in scalable quantum networks is demonstrated by these results.

 BQT-MDQW using the W-Bell state shows a significant difference from unidirectional methods in terms of memory utilization as illustrated in Figure \ref{fig5}(f). There is an obvious correlation between network complexity and resource usage, as memory utilization increases with the number of nodes. The plot of the results demonstrates that bidirectional routing is more effective in reducing resource cost than unidirectional teleportation, which continually uses more memory. As the network grows, the difference becomes more obvious, indicating that bidirectional quantum teleportation might be a better approach for large-scale quantum networks. In dynamic entangled channels, the W-Bell state improves performance by facilitating effective routing while preserving a fair trade-off between resource usage and fidelity. Memory utilization follows a similar trend ($0.0157$ vs. $0.0137$) and the forward path shows higher CPU utilization $(0.68)$ than the backward path $(0.63)$ in bigger networks, such as size \textbf{ 100}. However, in a network of size \textbf{150}, even though the path lengths are the same (8 nodes), the forward path uses significantly more CPU $(0.67)$ and memory $(0.0119)$ than the reverse path $(0.23 and 0.0052)$. This suggests that quantum state routing efficiency may have asymmetries. Despite both paths traversing 24 nodes, the forward path demonstrates higher CPU use (0.39) and memory utilization (0.0198) in the largest network studied \textbf{ (size 200)} than the reverse path $(0.25 and 0.0161)$.\\

{\bf GHZ-Bell state:} The circuit-based performance of GHZ-Bell states about the number of nodes in the quantum network is examined in Figure \ref{fig:}. The throughput of the quantum teleportation network is shown in Figure \ref{fig:}(a) as the number of nodes increases. The expected quantity of end-to-end qubits that are effectively generated is known as throughput. In the quantum network, a higher throughput indicates a more effective communication channel. Figure \ref{fig7}(a) illustrates how unidirectional teleportation offers more optimal routing options than bidirectional teleportation, the former typically achieves higher throughput. The accuracy of quantum state transfer throughout the network is the main topic of Figure \ref{fig7}(b). The precision and caliber of the entangled connections used in the teleportation process are measured by fidelity. The graph shows that as compared to bidirectional quantum teleportation, unidirectional teleportation consistently delivers superior fidelity. The difference is explained by the fact that unidirectional systems choose the most reliable entangled couples, increasing the total success rate of transmissions of quantum states. The memory utilization of the BQT protocol is examined in Figure \ref{fig7}(c). Memory utilization is the ratio of entangled pairs used to all pairs in the network. According to the results, unidirectional teleportation uses more memory than bidirectional teleportation. This can lead to memory shortages and increased decoherence risks since bidirectional quantum teleportation needs two simultaneously entangled links, which increases memory consumption and keeps track of periods for waiting connections.\\

{\bf Cluster-Bell state:} In Figure \ref{fig8}, the circuit-based performance of Cluster-Bell states in a quantum network. As the number of nodes in the quantum network increases, each subfigure focuses on a different performance parameter. The throughput of the quantum teleportation network using Cluster-Bell states is shown in Figure \ref{fig8}(a). The effective rate of successful quantum state transfers over a network is measured by throughput. The throughput typically rises as the number of nodes increases, demonstrating the network's robustness in effectively managing quantum communications. The results demonstrate optimal throughput efficiency compared to different entangled states, particularly in unidirectional teleportation scenarios, using Cluster-Bell states, which offer more flexible entanglement arrangements. The fidelity of quantum state transmission with Cluster-Bell states is assessed in Figure \ref{fig8}(b). By measuring how closely the received states resemble the original ones, fidelity quantifies the accuracy and reliability of the transmitted states. The outcomes indicate how well Cluster-Bell states preserve quantum information during teleportation procedures, with fidelity remaining good across a range of node counts. As the network grows, the fidelity also shows a minor performance increase over other state types, demonstrating how Cluster-Bell states can successfully preserve high-quality state transfers even in the face of multi-hop communications' complexity. The Figure \ref{fig8}(c) uses Cluster-Bell states to evaluate memory utilization in the context of circuit-based teleportation. The ratio of used entangled pairs to all pairs available in the quantum network is used to compute the amount of memory used. The memory utilization improves as the number of nodes increases, indicating that Cluster-Bell states are being used to effectively use available resources. The results make it clear that because of its basic features that facilitate effective entanglement distribution among nodes, this kind of entangled state functions with lower memory overhead in a multi-hop situation than other states.

\subsubsection{Comparative Analysis of Entangled State Selection in Quantum Networks and Simulation Frameworks:} 

The efficiency of the network protocol is essential for the successful transfer of information across numerous nodes in the growing field of quantum communications. Figures 5 and 6 illustrate the two different simulation methodologies that are compared in this work. In particular, simulations within the quantum network framework SimQN, where a quantum network is created using the Waxman model, are the subject of Figure \ref{fig5}. Using the W-Bell state as the entangled resource for teleportation protocols, this framework makes it easier to analyze network performance measures including throughput, fidelity, and memory utilization. 

However, Figure \ref{fig7} describes the quantum simulation method that uses the Quantum Dijkstra algorithm along with the Quantum QASM simulator and Quatum walk. This approach improves the study of quantum network shortest path calculations by showing the performance improvements made possible by the use of quantum walks and optimized routing. The W-Bell state is still the major entangled resource used in the simulations, giving a uniform framework for comparison between the two models. This study aims to show the advantages and disadvantages of both simulation approaches with regard to their ability to properly simulate and improve quantum network communications by comparing the results from Quantum Dijkstra algorithm, Quantum Walk and the QASM simulator with those from the implementation of SimQN. The analysis of the W-Bell state in each setting highlights how important it is for improving quantum teleportation techniques, which advances our knowledge of how it helps enable secure communication in quantum networks.

Studying quantum communication with teleportation methods successfully demonstrates the advantages and difficulties of different entangled states in a range of simulation environments. The performance characteristics of several simulation techniques are compared in this work, as shown in Figures \ref{fig5}, \ref{fig7}, \ref{fig8}. While in Figure \ref{fig5} describes the quantum simulation performance using the Quantum QASM simulator, Quantum Walk, and the Quantum Dijkstra algorithm, Figure \ref{fig5} presents the classical simulation results for W-Bell states using the SimQN network framework based on the Waxman model. 

The superiority of the $W-Bell$ state over the $GHZ-Bell$ and Cluster-Bell states for BQT-MDQW is demonstrated by the comparison of Figures \ref{fig5}, \ref{fig7}, \ref{fig8}. While the $GHZ-Bell$ and $Cluster-Bell$ states impose limitations that limit transmission rates, the W-Bell state achieves higher throughput by enabling more efficient routing. While $GHZ-Bell$ states are more likely to decohere, $Cluster-Bell$ states do slightly better but still fall short, while the W-Bell state maintains higher accuracy even as network size increases. In addition, the W-Bell state uses less resource than the other two states since it requires less entangled links, which maximizes memory utilization. Also, by improving error reduction and increasing its resistance to decoherence, its entanglement structure guarantees sustained quantum communication. In quantum teleportation networks, the W-Bell state is the best option overall because of its higher efficiency, security, and adaptability.

\section{Conclusion}
The bidirectional quantum teleportation protocol is completed with the successful transfer of Alice's and Bob's quantum states across the network. The W-Bell state, the GHZ-Bell state, and the Cluster-Bell state serve as the entangled resource. In summary, this work demonstrates how quantum computing could solve graph algorithms and optimization issues, especially when used with the Quantum Dijkstra Algorithm and Quantum Walk, which drastically reduces time complexity to $O(log(N)^2)$ from the traditional $O(N^2)$. By simulating 200 quantum nodes over a wide region, the SimQN framework easily models networks using the Waxman model. Error rates, circuit optimization, and scalability issues still need to be addressed despite encouraging outcomes. Performance measures including memory utilization, fidelity, and throughput show how various entangled states affect the effectiveness of quantum networks, with bidirectional quantum teleportation working better in larger networks. These results point to a fundamental change in computational methods by indicating that quantum computing provides quicker and more effective solutions to challenging graph issues. Future studies will concentrate on scalable quantum network systems, circuit optimization, and error reduction. In the end, this study demonstrates how quantum computing can transform shortest-path issues as well as more general optimization challenges, with important implications for network structure, quantum communication, and cryptography.

In the future, we aim to integrate unidirectional and BQT-MDQW to establish fault-tolerant quantum communication. This integration will facilitate dynamic entanglement routing, optimizing the entanglement distribution across large-scale quantum networks. Furthermore, self-correcting mechanisms that take advantage of quantum error correction and entanglement purification will enhance network resilience against decoherence and transmission errors. By combining these techniques, we can ensure a robust and high-fidelity quantum information transfer infrastructure, paving the way for a scalable and efficient global quantum internet capable of supporting secure communication, distributed quantum computing, and advanced cryptographic protocols.

\section*{Declaration}
{\bf Acknowledgments:} Princess Nourah bint Abdulrahman University Researchers Supporting Project number (PNURSP2025R752),Princess Nourah bint Abdulrahman University, Riyadh, Saudi Arabia . This paper also derived from a research grant funded by the Research, Development, and Innovation Authority (RDIA) - Kingdom of Saudi Arabia - with grant number (13325-psu-2023-PSNU-R-3-1-EF). Additionally, the authors would like to thank Prince Sultan University for their support.\par

{\bf Disclosures:} The authors declare that they have no known competing financial interests or personal relationships that could have appeared to influence the work reported in this paper.\par 

{\bf Data availability:} The datasets used and/or analysed during the current study available from the corresponding author on reasonable request.

\section*{Appendices}
\appendix

\section{Bidirectional Quantum Teleportation using a GHZ-Bell State:}\label{12}
To begin, we assume that Alice and Bob share an entangled composite state, referred to as the GHZ-Bell state, which is given by:

\begin{multline}
    |\phi\rangle_{12345} = \frac{1}{\sqrt{2}} \left( |000\rangle + |111\rangle \right)_{123} \otimes \frac{1}{\sqrt{2}} \left( |00\rangle + |11\rangle \right)_{45}\\ 
    = \frac{1}{\sqrt{2}} \left( |00000\rangle + |00011\rangle + |11100\rangle + |11111\rangle \right)_{12345}.
\end{multline}

In this state, Bob holds particles 2, 3, and 4, while Alice holds particles 1 and 5. This bipartite distribution of subsystems serves as the basis for the quantum communication of the two parties and reflects their spatial separation.

\begin{figure*}[!t]
    \centering
    \includegraphics[width=1\textwidth]{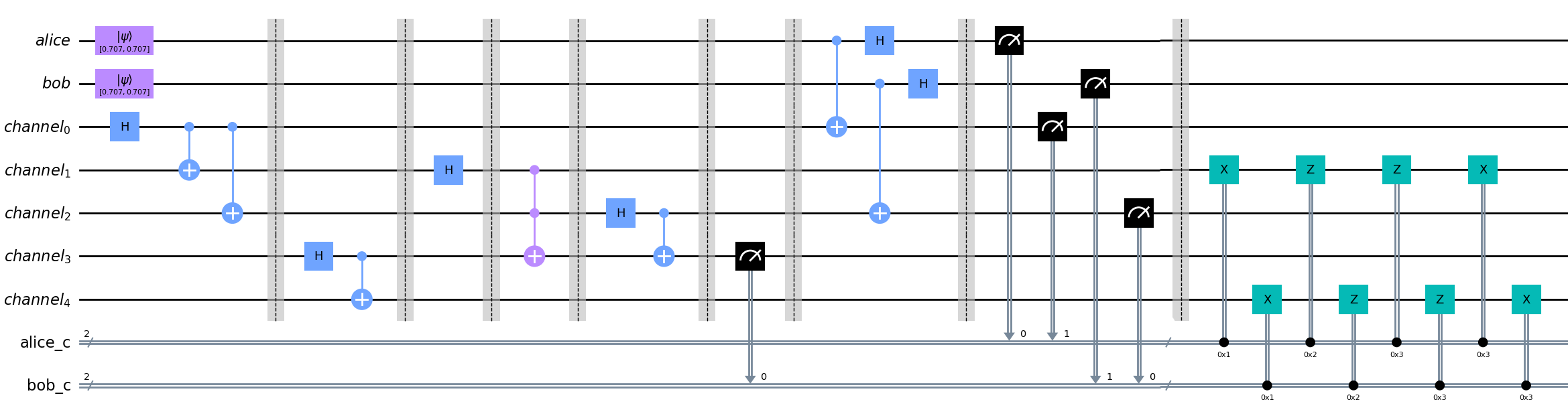}
    \caption{Quantum circuit for bidirectional quantum teleportation using a GHZ-Bell state as the shared entangled resource.}
    \label{fig9}
\end{figure*}

The initial state of the entire system, including Alice's and Bob's local qubits, can then be expressed as:

\begin{align}
    |\Phi\rangle_{12345AB} = |\phi\rangle_{12345} \otimes |\phi\rangle_A \otimes |\phi\rangle_B,
\end{align}

where $|\phi\rangle_A$ and $|\phi\rangle_B$ represent the unknown quantum states of Alice's and Bob's qubits, respectively. As shown in the figure \ref{fig9}, this total system state forms the starting point for analyzing entanglement-based quantum communication protocols.

The circuit shown in \ref{fig9} for bidirectional quantum teleportation is designed to use entanglement as a resource distributed across multiple nodes. Entanglement is a resource that is dispersed over several nodes in the circuit for bidirectional quantum teleportation. A Hadamard gate and Controlled-NOT (CNOT) gates are used to construct the GHZ state, $\frac{1}{\sqrt{2}}(|000\rangle + |111\rangle)$, which ensures maximal entanglement for the teleportation protocol. Additional qubits are created in a Bell state, $\frac{1}{\sqrt{2}}(|00\rangle + |11\rangle)$, and entangled with the GHZ state using additional CNOT operations to try to allow multiparty entanglement. Alice and Bob then distribute these entangled qubits across the network. Both parties collapse the entangled state and project the remaining qubits into a correlated state by performing local Bell-State Measurements (BSM) on their respective particles and unknown quantum states, $|\phi\rangle_A$ and $|\phi\rangle_B$. Since the measurement results are sent across classical communication channels, the original quantum states on the corresponding target qubits can be recovered by applying unitary corrections using Pauli operators (gates X and Z). The successful transfer of quantum states across numerous nodes is ensured by applying this procedure continuously at each network hop.

To perform the calculations step by step as described, we proceed as follows:

\section{Calculation Steps for Bidirectional Quantum Teleportation}

\emph{\textbf{Step 1: Applying the Hadamard Gate to qubit 2.}} 
Bob uses a Hadamard gate (H) on the GHZ state's second qubit. The state changes in the following ways:
\begin{multline}
    |\phi\rangle_{12345} = \frac{1}{\sqrt{2}} \big(|000\rangle + |111\rangle\big)_{123} \otimes \frac{1}{\sqrt{2}} \big(|00\rangle + |11\rangle\big)_{45} \\
    \xrightarrow{H_2} \frac{1}{2\sqrt{2}} \big[|00\rangle_{13}(|0\rangle + |1\rangle)_2 + |11\rangle_{13}(|0\rangle - |1\rangle)_2\big] \\
    \otimes \big(|00\rangle + |11\rangle\big)_{45}.
\end{multline}

\emph{\textbf{Step 2: Applying the Controlled-Controlled-NOT (CCNOT) Gate.}} 
Using qubit 2 as a control qubit, another control qubit in state $|1\rangle$, and a target qubit in state $|0\rangle$, Bob applies a CCNOT gate to qubit 2. The target qubit stays the same if qubit 2 is in the state $|0\rangle$. The target qubit flips if qubit 2 is in the state $|1\rangle$. Qubit 2 stays in the state $|0\rangle$ if the target qubit measurement returns $|0\rangle$. The state turns into:
\begin{multline}
    |\phi'\rangle_{12345} \xrightarrow{CCNOT} \frac{1}{4} \big[|00000\rangle + |01000\rangle + |10100\rangle -\\
    |11110\rangle + |00011\rangle + |01011\rangle + |10111\rangle + |11100\rangle\big]_{12345}
\end{multline}

\emph{\textbf{Step 3: Applying the Hadamard and CNOT gate.}} 
Qubit 3 is the control qubit, while particle 4 is the target qubit. Bob applies a Hadamard to qubit 3 and a CNOT gate to qubits 3 and 4. The status changes as follows:
\begin{multline}
    |\phi"\rangle_{12345} \xrightarrow{H_{3}} \\ \frac{1}{4\sqrt{2}} \big[|0000\rangle_{1245}(|0\rangle + |1\rangle)_{3} + |0100\rangle_{1245}(|0\rangle + |1\rangle)_{3} \\ + |1000\rangle_{1245}(|0\rangle - |1\rangle)_{3} - |1110\rangle_{1245}(|0\rangle - |1\rangle)_{3} \\ +  |0011\rangle_{1245}(|0\rangle + |1\rangle)_{3} + |0111\rangle_{1245}(|0\rangle + |1\rangle)_{3} \\ +  |1011\rangle_{1245}(|0\rangle - |1\rangle)_{3} - |1100\rangle_{1245}(|0\rangle - |1\rangle)_{3}\big].
\end{multline}

We apply the CNOT gate:
\begin{multline}
    |\phi'''\rangle_{12345} \xrightarrow{CNOT_{3,4}} \\ \frac{1}{8} \big[|00000\rangle_{12345} + |00110\rangle_{12345} + |01000\rangle_{12345} \\ + |01110\rangle_{12345}+|10000\rangle_{12345} - |10110\rangle_{12345} \\ -|11010\rangle_{12345}+|11100\rangle_{12345}+|00011\rangle_{12345} \\ + |00101\rangle_{12345}+|01011\rangle_{12345}+|01101\rangle_{12345} \\ + |10011\rangle_{12345}-|10101\rangle_{12345} \\ -|11000\rangle_{12345}+|11110\rangle_{12345}\big].
\end{multline}

\emph{\textbf{Step 4: Measurement of Qubit 4.}} 
Bob uses the computational basis to do a measurement on qubit 4. The state of the system collapses in correlation with the measurement's result, which might be either $|0\rangle$ or $|1\rangle$. 

If qubit 4 is measured to be in state $|0\rangle$, the state of the system becomes:
\begin{multline}
    |\phi_{\text{measured}}\rangle_{12345} = \frac{1}{4} \big[|0000\rangle_{1235} + |0011\rangle_{1235} \\ + |0100\rangle_{1235} + |0111\rangle_{1235} 
    + |1000\rangle_{1235} \\ - |1011\rangle_{1235} - |1101\rangle_{1235} + |1110\rangle_{1235}\big].
\end{multline}

The system's state is as follows if qubit 4 is measured to be in state $|1\rangle$:
\begin{multline}
    |\phi_{\text{measured}}\rangle_{12345} = \frac{1}{4} \big[|0001\rangle_{1235} + |0010\rangle_{1235} + |0101\rangle_{1235} \\ + |0110\rangle_{1235} 
    + |1001\rangle_{1235} - |1010\rangle_{1235} \\ - |1100\rangle_{1235} + |1111\rangle_{1235}\big].
\end{multline}

\emph{\textbf{Step 5: Applying CNOT Gates to Alice's and Bob's States.}} 
Alice's initial state is:
\begin{align}
    |\phi\rangle_A = \alpha |0\rangle_A + \beta |1\rangle_A,
\end{align}
and Bob's initial state is:
\begin{align}
    |\phi\rangle_B = \gamma |0\rangle_B + \eta |1\rangle_B.
\end{align}

Bob's qubit is the target of the CNOT gate, while Alice's qubit serves as the control qubit. Alice and Bob's combined beginning state is:
\begin{align}
    |\phi\rangle_{AB} = \left(\alpha |0\rangle_A + \beta |1\rangle_A\right) \otimes \left(\gamma |0\rangle_B + \eta |1\rangle_B\right).
\end{align}

Expanding this state:
\begin{align}
    |\phi\rangle_{AB} = \alpha\gamma |00\rangle_{AB} + \alpha\eta |01\rangle_{AB} + \beta\gamma |10\rangle_{AB} + \beta\eta |11\rangle_{AB}.
\end{align}

When the CNOT transformations are applied, the state following the CNOT gate is:
\begin{align}
    |\phi'\rangle_{AB} = \alpha\gamma |00\rangle_{AB} + \alpha\eta |01\rangle_{AB} + \beta\gamma |11\rangle_{AB} + \beta\eta |10\rangle_{AB}.
\end{align}

After terms are modified, and the outcome is:
\begin{align}
    |\phi'\rangle_{AB} = \alpha\gamma |00\rangle_{AB} + \alpha\eta |01\rangle_{AB} + \beta\eta |10\rangle_{AB} + \beta\gamma |11\rangle_{AB}.
\end{align}

\emph{\textbf{Step 6: Applying Hadamard Gates.}}
%
Before the Hadamard gates were applied, the system was the following:
\begin{align}
    |\phi'\rangle_{AB} = \alpha\gamma |00\rangle_{AB} + \alpha\eta |01\rangle_{AB} + \beta\eta |10\rangle_{AB} + \beta\gamma |11\rangle_{AB}.
\end{align}

The Hadamard gate acts on a single qubit as follows:
\begin{align}
    H|0\rangle = \frac{1}{\sqrt{2}}(|0\rangle + |1\rangle), \quad H|1\rangle = \frac{1}{\sqrt{2}}(|0\rangle - |1\rangle).
\end{align}

Applying the Hadamard gate to both Alice's qubit (A) and Bob's qubit (B), we calculate the transformations of each base state:
\begin{align}
    H_A H_B |00\rangle_{AB} &= \frac{1}{2}(|0\rangle_A + |1\rangle_A) \otimes (|0\rangle_B + |1\rangle_B), \\
    H_A H_B |01\rangle_{AB} &= \frac{1}{2}(|0\rangle_A + |1\rangle_A) \otimes (|0\rangle_B - |1\rangle_B), \\
    H_A H_B |10\rangle_{AB} &= \frac{1}{2}(|0\rangle_A - |1\rangle_A) \otimes (|0\rangle_B + |1\rangle_B), \\
    H_A H_B |11\rangle_{AB} &= \frac{1}{2}(|0\rangle_A - |1\rangle_A) \otimes (|0\rangle_B - |1\rangle_B).
\end{align}

Returning to the initial state after performing these changes:
\begin{align}
    |\phi''\rangle_{AB} &= \alpha\gamma \frac{1}{2}(|0\rangle_A + |1\rangle_A)(|0\rangle_B + |1\rangle_B) \nonumber \\
    &\quad + \alpha\eta \frac{1}{2}(|0\rangle_A + |1\rangle_A)(|0\rangle_B - |1\rangle_B) \nonumber \\
    &\quad + \beta\eta \frac{1}{2}(|0\rangle_A - |1\rangle_A)(|0\rangle_B + |1\rangle_B) \nonumber \\
    &\quad + \beta\gamma \frac{1}{2}(|0\rangle_A - |1\rangle_A)(|0\rangle_B - |1\rangle_B).
\end{align}

Combining and extending terms:
\begin{align}
    |\phi''\rangle_{AB} &= \frac{1}{2} \bigg[
        (\alpha\gamma + \alpha\eta + \beta\eta + \beta\gamma)|00\rangle_{AB} \nonumber \\
        &\quad + (\alpha\gamma - \alpha\eta + \beta\eta - \beta\gamma)|01\rangle_{AB} \nonumber \\
        &\quad + (\alpha\gamma + \alpha\eta - \beta\eta - \beta\gamma)|10\rangle_{AB} \nonumber \\
        &\quad + (\alpha\gamma - \alpha\eta - \beta\eta + \beta\gamma)|11\rangle_{AB}
    \bigg].
\end{align}

\emph{\textbf{Step 7: Measurement and recovery.}} 
Bob and Alice use a classical channel to exchange the results of their measurements on their qubits. The original states $|\phi\rangle_A$ and $|\phi\rangle_B$ at the target locations are recovered by applying unitary adjustments (Pauli gates X and Z) based on the results.


\begin{thebibliography}{9}

\bibitem{1993}Charles H. Bennett, Gilles Brassard, Claude Cr´epeau, Richard Jozsa, Asher Peres, and William K.Wootters. Teleporting an unknown quantum state via dual classical and einstein-podolsky-rosen channels. Phys. Rev. Lett., 70:1895–1899, Mar 1993
\bibitem{1997}Dik Bouwmeester, Jian-Wei Pan, Klaus Mattle, Manfred Eibl, Harald Weinfurter, and Anton Zeilinger.Experimental quantum teleportation. Nature, 390(6660):575–579, Dec 1997.
\bibitem{2008}ARTI CHAMOLI and C. M. BHANDARI. Teleportation of unknown state by qutrits. International Journal of Quantum Information, 06(02):369–378, 2008.
\bibitem{24}N. C. Randeep, C. Anukrishna, and A. K. Neha Raj. Quantum teleportation scheme using entangled two ququads and its noise effects. Quantum Information Processing, 23(5):192, May 2024.
\bibitem{8}Xin-Wei Zha, Zhi-Chun Zou, Jian-Xia Qi, and Hai-Yang Song. Bidirectional quantum controlled teleportation via five-qubit cluster state. International Journal of Theoretical Physics, 52(6):1740–1744, Jun 2013
\bibitem{9}An Yan. Bidirectional controlled teleportation via six-qubit cluster state. International Journal of
Theoretical Physics, 52(11):3870–3873, Nov 2013
\bibitem{dijk}Dijkstra, E. W. (2022). A note on two problems in connexion with graphs. In Edsger Wybe Dijkstra: his life, work, and legacy (pp. 287-290).
\bibitem{Zurek}Zurek, Wojciech Hubert. "Decoherence, einselection, and the quantum origins of the classical." Reviews of modern physics 75.3 (2003): 715.
\bibitem{Chen}Chen, Xie, Zheng-Cheng Gu, and Xiao-Gang Wen. "Local unitary transformation, long-range quantum entanglement, wave function renormalization, and topological order." Physical Review B—Condensed Matter and Materials Physics 82.15 (2010): 155138.
\bibitem{Elliott}Elliott, Chip, David Pearson, and Gregory Troxel. "Quantum cryptography in practice." Proceedings of the 2003 conference on Applications, technologies, architectures, and protocols for computer communications. 2003.
\bibitem{10}Yu Wang, Yun Shang, and Peng Xue. Generalized teleportation by quantum walks. Quantum Information Processing, 16(9):221, Jul 2017.
\bibitem{BQT}Ikken, N., Slaoui, A., Ahl Laamara, R., and Drissi, L. B. (2023). Bidirectional quantum teleportation of even and odd coherent states through the multipartite Glauber coherent state: Theory and implementation. Quantum Information Processing, 22(10), 391.
\bibitem{11}Salvador El´ıas Venegas-Andraca. Quantum walks: a comprehensive review. Quantum Information Processing, 11(5):1015–1106, Oct 2012.
\bibitem{12}Andris Ambainis, Eric Bach, Ashwin Nayak, Ashvin Vishwanath, and John Watrous. One-dimensional quantum walks. In Proceedings of the Thirty-Third Annual ACM Symposium on Theory of Computing, STOC ’01, page 37–49, New York, NY, USA, 2001. Association for Computing Machinery
\bibitem{13}Karuna Kadian, Sunita Garhwal, and Ajay Kumar. Quantum walk and its application domains: A systematic review. Computer Science Review, 41:100419, 2021.
\bibitem{14}Xiaogang Qiang, Shixin Ma, and Haijing Song. Review on quantum walk computing: Theory, implementation, and application, 2024.
\bibitem{15}Heng-Ji Li, Xiu-Bo Chen, Ya-Lan Wang, Yan-Yan Hou, and Jian Li. A new kind of flexible quantum teleportation of an arbitrary multi-qubit state by multi-walker quantum walks. Quantum Information Processing, 18(9):266, Jul 2019
\bibitem{16}Xiaoxiao Chen and Xiaoping Lou. An efficient verifiable quantum secret sharing scheme via quantum walk teleportation. International Journal of Theoretical Physics, 61(4):99, Apr 2022.
\bibitem{Yurke}Yurke, Bernard, and John S. Denker. "Quantum network theory." Physical review A 29.3 (1984): 1419.
\bibitem{maj} Slaoui, A., Ikken, N., Btissam Drissi, L., Ahl Laamara, R. (2023). Quantum communication protocols: From theory to implementation in the quantum computer. IntechOpen. doi: 10.5772/intechopen.1002792
\bibitem{min}El Kirdi, M., Slaoui, A., Ikken, N., Daoud, M., Laamara, R. A. (2023). Controlled quantum teleportation between discrete and continuous physical systems. Physica Scripta, 98(2), 025101.
\bibitem{bet}Dakir, Y., Slaoui, A., Mohamed, AB.A. et al. Quantum teleportation and dynamics of quantum coherence and metrological non-classical correlations for open two-qubit systems. Sci Rep 13, 20526 (2023).
\bibitem{een}Slaoui, A., El Kirdi, M., Ahl Laamara, R. et al. Cyclic quantum teleportation of two-qubit entangled states by using six-qubit cluster state and six-qubit entangled state. Sci Rep 14, 15856 (2024).
\bibitem{17}Kempe, J. (2003). Quantum random walks: an introductory overview. Contemporary Physics, 44(4), 307–327.
\bibitem{18}Liu, Y., Koehler, G. J. (2010). Using modifications to Grover’s Search algorithm for quantum global optimization. European Journal of Operational Research, 207(2), 620-632.
\bibitem{Winterbone2015} Bennett, C.H., Brassard, G., Crépeau, C., Jozsa, R., Peres, A., Wootters, W.K.: Teleporting an
unknown quantum state via dual classical and Einstein-Podolsky-Rosen channels. Phys. Rev. Let. 70, 1895–1899 (1993)
\bibitem{2013} Yu, X.-T., Xu, J., Zhang, Z.-C.: Distributed wireless quantum communication networks. Chin. Phys. B 22(9), 090311 (2013)
\bibitem{2014}Wang, K., Yu, X.-T., Lu, S.-L., Gong, X.-Y.: Quantum wireless multihop communication based on arbitrary Bell pairs and teleportation. Phys. Rev. A 89(2), 022329 (2014)
\bibitem{2006}Agrawal, P., Pati, A.: Perfect teleportation and superdense coding with W states. Phys. Rev. A 74, 062320(2006)
\bibitem{2003}J. Kempe, “Quantum random walks: an introductory overview,” Contemporary Physics, vol. 44, no. 4, pp. 307–327, 2003.
\bibitem{2022}Dijkstra, E. W. (2022). A note on two problems in connexion with graphs. In Edsger Wybe Dijkstra: his life, work, and legacy (pp. 287-290).
\bibitem{2024}Tsoufis, C. (2024). Quantum computing: algorithms, implementations, and applications to cryptography and graph theory.
\bibitem{2020}Liu, Y., Koehler, G. J. (2010). Using modifications to Grover’s Search algorithm for quantum global optimization. European Journal of Operational Research, 207(2), 620-632.
\bibitem{2023}Chen, Lutong, et al. "SimQN: A network-layer simulator for the quantum network investigation." IEEE Network 37.5 (2023): 182-189.
\bibitem{huang2022socially}Huang, Shao-Min, et al. "Socially-aware concurrent entanglement routing with path decomposition in quantum networks." GLOBECOM 2022-2022 IEEE Global Communications Conference. IEEE, 2022.
\bibitem{1998routing} Waxman, Bernard M. "Routing of multipoint connections." IEEE journal on selected areas in communications 6.9 (1988): 1617-1622.
\bibitem{yin2012quantum}Yin, Juan, et al. "Quantum teleportation and entanglement distribution over 100-kilometre free-space channels." Nature 488.7410 (2012): 185-188.
\bibitem{lee2000entanglement}Lee, Jinhyoung, and M. S. Kim. "Entanglement teleportation via Werner states." Physical review letters 84.18 (2000): 4236.
\end{thebibliography}
\end{document}